%% file: main.tex
\documentclass[trackchanges,twocolumn]{aastex7}
\graphicspath{{./}{figures/}}

\usepackage{booktabs} 

\usepackage{CJKutf8} 

\usepackage[acronym,nohypertypes={acronym}]{glossaries} 
\loadglsentries{glossary}

\newcommand{\glsplwithcite}[2]{%
    \ifglsused{#1}{\glspl{#1} (#2)}{\glsentrylongpl{#1} (\glsentryshortpl{#1}; #2)\glsunset{#1}}
}

\newcommand{\todo}[1]{{#1}}

\shorttitle{An Autoencoder Framework for Binary Identification Applied to 47 Tucanae}
\shortauthors{G\'eron et al.}

\begin{document}


\title{Unresolved Binary Systems in the Rubin Era I: An Autoencoder Framework for Binary Identification Applied to 47 Tucanae}

\correspondingauthor{Tobias G\'eron}
\email{tobias.geron@utoronto.ca}

\author[orcid=0000-0002-6851-9613,gname=Tobias,sname=G\'eron]{Tobias G\'eron}
\affiliation{Dunlap Institute for Astronomy \& Astrophysics, University of Toronto, 50 St. George Street, Toronto, ON M5S 3H4, Canada}
\email{tobias.geron@utoronto.ca} 

\author[0000-0002-5522-0217]{Alexander Laroche}
\affiliation{Dunlap Institute for Astronomy \& Astrophysics, University of Toronto, 50 St. George Street, Toronto, ON M5S 3H4, Canada}
\affiliation{David A. Dunlap Department of Astronomy \& Astrophysics, University of Toronto, 50 St. George Street, Toronto, ON M5S 3H4, Canada}
\email{alex.laroche@mail.utoronto.ca}

\author[0000-0003-2573-9832]{Joshua S. Speagle \begin{CJK*}{UTF8}{gbsn}(沈佳士)\end{CJK*}}
\affiliation{David A. Dunlap Department of Astronomy \& Astrophysics, University of Toronto, 50 St. George Street, Toronto, ON M5S 3H4, Canada}
\affiliation{Department of Statistical Sciences, University of Toronto, 9th Floor, 700 University Avenue, Toronto, ON M5S 3G3, Canada}
\affiliation{Dunlap Institute for Astronomy \& Astrophysics, University of Toronto, 50 St. George Street, Toronto, ON M5S 3H4, Canada}
\affiliation{Data Sciences Institute, University of Toronto, 10th Floor, 700 University Avenue, Toronto, ON M7A 2S4, Canada}
\email{j.speagle@utoronto.ca}

\author[0000-0001-7081-0082]{Maria R. Drout}
\affiliation{David A. Dunlap Department of Astronomy \& Astrophysics, University of Toronto, 50 St. George Street, Toronto, ON M5S 3H4, Canada}
\email{-}



\begin{abstract}
Binary systems are extremely common and influence many areas of astrophysics, such as stellar evolution and cluster dynamics. In this work, we introduce \texttt{flexAE}, a framework that combines a classical autoencoder architecture with a dedicated classifier component. It is designed to distinguish between single stars and binary systems based on their broadband spectral energy distributions, using all available photometry simultaneously while accounting for observational uncertainties. We demonstrate the capability of \texttt{flexAE} by using it to identify unresolved main sequence binaries in the outskirts of the 47 Tucanae globular cluster (NGC 104) with photometric data from the Vera C. Rubin Observatory. We train the model on a simulated sample of single star and binary systems created with stellar atmosphere models. The model accurately reconstructs the photometric input features and achieves a classification accuracy of \todo{0.85} on the simulated test set. We then apply the trained model to a sample of \todo{1,424} cluster members found in Rubin DP1 with reliable $gri$ photometry, of which \todo{32} are identified by the model as unresolved main sequence binaries. This corresponds to an observed binary fraction of \todo{$2.2^{+0.5}_{-0.3}\%$}. We estimate that the intrinsic main sequence binary fraction lies between \todo{$3.7^{+0.8}_{-0.5}\%$} and \todo{$7.5^{+1.5}_{-1.1}$}\%. The binary fraction stays constant between \todo{$18-36$} arcmin (\todo{$5.7-11.4$} half-light radii) from the cluster center. We highlight that \texttt{flexAE} can be easily adapted for other use cases. This makes \texttt{flexAE} well-suited for large-scale binary classification in upcoming wide-field surveys, such as Rubin LSST. The code is publicly available at \url{https://github.com/tobiasgeron/flexAE}.
\end{abstract}




\keywords{\uat{Binary stars}{154}; \uat{Stellar photometry}{1620}; \uat{Main sequence stars}{1000}; \uat{Stellar populations}{1622}; \uat{Star clusters}{1567}; \uat{Astronomy software}{1855}; \uat{Classification}{1907}}


\section{Introduction}
\label{sec:introduction}

Binary star systems are very common; we know that more than half of massive ($M_* \gtrsim 8 M_\odot$) stars have a binary companion \citep{sana_2012, sana_2013,sana_2025, rizzuto_2013, kobulnicky_2014,offner_2023}. It is also well established that binary systems are crucial to multiple areas of astrophysics. For example, they are the precursors to \glsplwithcite{SN Type Ia}{\citealp{whelan_1973,maoz_2014,ruiter_2025}} and gravitational wave sources \citep{belczynski_2002,postnov_2014,broekgaarden_2022}. However, despite their importance and prevalence, many questions about their formation, evolution, and interaction remain. Some outstanding questions include the physics of binary mass transfer (e.g., see \citealp{picco_2026}), the outcomes of common envelope evolution \citep{ivanova_2013}, and the physics of SN kicks \citep{bray_2016, giacobbo_2020}. These open questions cause further downstream uncertainties in other fields of astrophysics, such as the amount of ionizing radiation coming from stripped stars with a binary companion that is often not accounted for in simulations \citep{stanway_2016, gotberg_2019,lecroq_2024}, the demographics of SN types \citep{ercolino_2026}, and the rates of compact object formation and mergers \citep{kalogera_2007,broekgaarden_2022,mandel_2022}. For a broader overview of the effect of stellar binarity on different fields in astrophysics, see \citet{breivik_2019} and references therein.

Binary systems also impact the dynamics and evolution of stellar clusters while providing important insights into their contents and formation. For example, they add energy to the cluster through binary interactions that support the cluster against collapse \citep{gao_1991, fregeau_2003} and drive mass segregation within the cluster \citep{heggie_2006, gill_2008, bianchini_2016b}. Binaries can also affect cluster dynamics by influencing the velocity dispersion of stars in the cluster \citep{bianchini_2016}, which in turn results in an overestimation of the cluster mass if not properly corrected for \citep{aros_2021}. Additionally, as noted by \citet{cordoni_2025}, not accounting for unresolved binaries will also cause errors in the mass and luminosity functions. Many interesting and exotic objects, such as millisecond pulsars, hypervelocity stars, cataclysmic variables, low-mass X-ray binaries, and blue straggler stars, could result from the large number of stellar interactions in the dense cluster environment (e.g., see \citealp{ye_2019,ye_2024,evans_2025,muller_horn_2025,grondin_2026}).

Furthermore, the radial profile of the binary fraction in clusters can be used to find \glspl{IMBH} \citep{aros_2021}. Interestingly, the binary fraction is significantly lower in globular clusters compared to field stars \citep{cool_2002, sollima_2007}, with a noticeable anti-correlation between the binary fraction and the mass of the cluster \citep{milone_2012}, as the dense cluster environment disrupts binary systems \citep{ivanova_2005}. The fraction of binary systems composed of two \gls{MS} stars in 47 Tucanae (NGC 104), a prominent globular cluster in the southern hemisphere, varies between $0.5 - 4.26$\% \citep{milone_2012, ji_2015, cordoni_2025, muller_horn_2025}, though the exact values depend on what mass ratios are considered and at what radius in the cluster it is measured.

Thus, binaries influence many areas of astrophysics, and the first step to any study of binary systems is reliably finding them. Multiple techniques have been used in the literature to identify binary systems, such as radial velocity variability analysis (e.g., see \citealt{latham_1996, sana_2013, jack_2019, villasenor_2021, sana_2025}). However, this technique is observationally expensive, as repeated spectroscopic observations are needed. It is also biased towards bright systems and binaries with short and eccentric orbits. Another technique uses the stellar photometric variability of binary systems (e.g., \citealt{kaluzny_1996,kirk_2016,soszynski_2016,mowlavi_2023}). However, this technique also requires repeated observations and is similarly biased towards short-period and eclipsing systems.

Many studies opt to use \gls{CMD}-based approaches instead (e.g., \citealt{rubenstein_1997, milone_2012, mohandasan_2024, cordoni_2025}), which leverage the fact that systems with flux contributions from multiple stars can exhibit unique combinations of magnitude and color, compared to single stars. For example, \gls{MS}+\gls{MS} binary systems will be slightly brighter and redder relative to single \gls{MS} stars (e.g., see \citealp{cordoni_2025}), while \gls{MS}+\gls{WD} and \gls{MS}+helium star binaries both lie blueward of the zero age main sequence (e.g., see \citealp{drout_2023, grondin_2024}). Notably, these \gls{CMD}-based methods are fast and observationally cheap, as they do not require repeat measurements, which allows for efficient analysis of statistically significant samples. In addition, the observational biases of \gls{CMD}-based methods are distinct from (and therefore complementary to) the aforementioned radial velocity and photometric variability methods. In particular, \gls{CMD}-based methods are not biased towards certain orbital configurations. However, they do require that the presence of the two sources is discernible from their combined \gls{SED}. This typically requires either a significant flux contribution from both stars or distinct stellar temperatures and broad wavelength coverage. For example, limits on photometric errors and differential reddening typically limit the detection of \gls{MS}+\gls{MS} binaries in stellar clusters to systems with high mass ratios ($q \gtrsim0.7$; e.g., see \citealt{mohandasan_2024,cordoni_2025}). 
Instead of using \glspl{CMD}, \citet{neugent_2021} used color-color diagrams to distinguish single \glspl{RSG} from \glspl{RSG} with an O- or B-type \gls{MS} companion by looking for excess blue or ultraviolet light. \citet{ogrady_2024} used a similar approach to look for \glspl{YSG} with an O- or B-type \gls{MS} companion.

More recently, different \gls{ML} models have also been successful at classifying binaries. For example, \citet{grondin_2024} trained a \gls{SVM} to look for \gls{WD}+\gls{MS} binaries using 10 different colors. The \gls{ALeRCE} broker for Rubin \gls{LSST} uses a balanced random forest to classify eclipsing binaries using variability metrics from the light curve from \gls{ZTF} combined with colors from AllWISE and \gls{ZTF} as input features \citep{sanchez_saez_2021}. 

However, while the aforementioned \gls{ML} models are very powerful, they are also typically tailored to identify specific binary populations or use specialized datasets. Furthermore, many of these techniques are also restrictive, which means that genuinely novel or unexpected events may go undetected. This highlights a gap in the field. There is a need for a classification technique that has all the advantages of \gls{CMD}-based methods (i.e., fast and cheap) that uses all available filters and colors, yet remains versatile enough to adapt to any type of stellar population. In this work, we present \texttt{flexAE}, a novel and flexible \gls{ML} approach to identifying binary stars that is based on a \gls{AE} framework that satisfies all criteria mentioned above. 

The need for such a framework has become particularly apparent with the recent commencement of the NSF-DOE Vera C. Rubin Observatory's \gls{LSST}. This survey will image $\sim$18,000 deg$^{2}$ in the southern sky up to a single-visit and co-added depth of $r \sim 24.0$ mag and $r \sim 26.9$ mag, respectively, in six photometric bands ($ugrizy$) with a targeted seeing of $\sim0.73$ arcsec in the $r$ band \citep{pstn_054, ivezic_2019}.\footnote{As a comparison, the DESI Legacy Imaging Surveys cover $\sim$14,000 deg$^{2}$ with three bands ($grz$) with an $r$-band depth of $\sim23.5$ \citep{dey_2019}.} With this footprint, depth, and photometric coverage, Rubin \gls{LSST} has the potential to identify an unprecedented number of binary systems if their broadband \glspl{SED} can be fully leveraged. The first preliminary data from the Rubin Observatory was released in \gls{DP1}, which included a field centered on the globular cluster 47 Tucanae \citep{rubin_dp1_2025}. This is the ideal opportunity to test \texttt{flexAE} and prepare for upcoming data releases from the Rubin Observatory, such as \gls{DP2} and \gls{DR1}, which will enable us to find even more binary systems in different stellar environments.

In this work, we introduce the \texttt{flexAE} framework and use it to find \gls{MS}+\gls{MS} binaries in 47 Tucanae using data from Rubin \gls{DP1}. However, we emphasize that \texttt{flexAE} is designed to be flexible and can easily be adapted to other use cases and work with data from other observatories with different photometric bands (e.g., we use \texttt{flexAE} in Appendix \ref{app:YSG} to identify binary systems with a \gls{YSG} and O- or B-type companion instead). The data that we use in this work is described in Section \ref{sec:data}. The \texttt{flexAE} framework, architecture, and training procedure are discussed in Section \ref{sec:ae}. The results are described and discussed in Section \ref{sec:results}. Finally, we summarize this work in Section \ref{sec:conclusion}. \texttt{flexAE} is publicly available on GitHub.\footnote{\url{https://github.com/tobiasgeron/flexAE}}

\section{Data}
\label{sec:data}

We use data from Rubin \gls{DP1}, which is discussed in Section \ref{sec:rubin_observatory}. We specifically use data from the 47 Tucanae field, which is described in more detail in Section \ref{sec:sample_selection}. Finally, we also use the stellar atmosphere models from \citet{castelli_2003}, which are discussed in Section \ref{sec:template_library}, to create a training set for the model.

\subsection{Vera C. Rubin Observatory}
\label{sec:rubin_observatory}

The Vera C. Rubin Observatory is located on Cerro Pach\'on, Chile. The observatory houses the Simonyi Survey Telescope, which will be used to conduct the planned 10-year \gls{LSST} \citep{ivezic_2019}. The deep-wide-fast survey mode will uniformly observe an area of 18,000 deg$^{2}$ up to a single-visit and co-added depth of $r \sim 24.0$ and $r \sim 26.9$, respectively \citep{pstn_054}, using six filters ($ugrizy$) with a wavelength range of 320 - 1050 nm. The primary and tertiary mirrors of the telescope (M1M3) are combined into a continuous surface with a diameter of 8.4m. The secondary mirror (M2) has a diameter of 3.42m \citep{ivezic_2019}.

In this work, we use the first set of data released by the Rubin Observatory, called \gls{DP1}. The observations for \gls{DP1} took place over a $\sim$1.5 month period between 24 October 2024 and 11 December 2024 \citep{rubin_dp1_2025, rubin_dp1_dataset}. The data in \gls{DP1} was obtained with a smaller version of the \gls{LSSTCam}, called \gls{LSSTComCam} \citep{rubin_comcam_2024}. Rubin \gls{DP1} consists of observations of $\sim15$ deg$^{2}$ in seven fields, including the 47 Tucanae Globular Cluster field, which we will focus on in this work. The 5$\sigma$ point source depth for the single-visit $g$-, $r$-, and $i$-band images in the 47 Tucanae field equals 24.14, 24.14, and 23.77 mag, respectively \citep{choi_2025}.

The data from Rubin \gls{DP1} were automatically reduced and analyzed by the Rubin Science pipelines \citep{rubin_pipelines_2025}. These pipelines perform tasks including (but not limited to) coaddition of images, background subtraction, photometric calibration, source detection, and source deblending \citep{rubin_pipelines_2025}. The final data products were accessed using the \gls{RSP} \citep{rubin_science_platform_2017, rubin_science_platform_2024}.

\subsection{47 Tucanae Sample Selection}
\label{sec:sample_selection}

\begin{figure*}
\includegraphics[width=\textwidth]{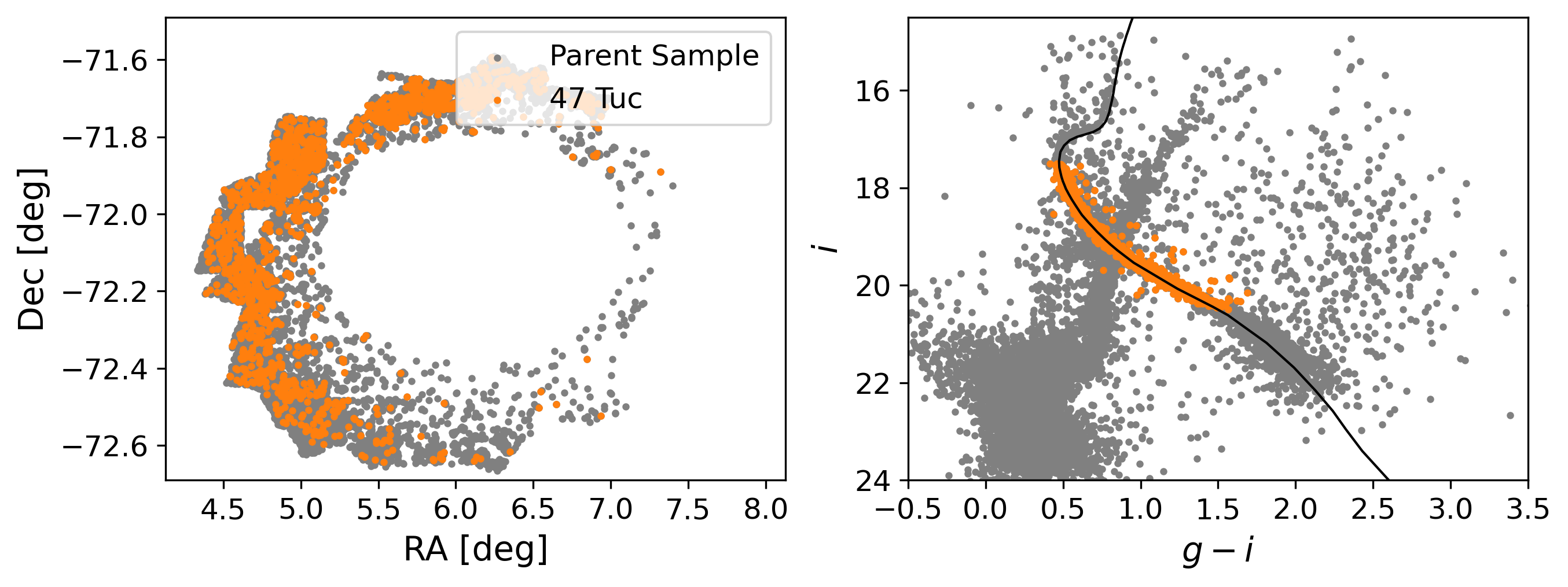}
    \caption{The left panel shows the location of all stars in the parent sample (grey) and the final 47 Tuc sample used in this work (orange). The right panel shows the $i$ vs. $g-i$ \gls{CMD}. The black line represents the PARSEC isochrone \citep{bressan_2012, chen_2015} assuming $d = 4.66$ kpc, $[M/H] = -0.5$, $E(B-V) = 0.025$, and $t_{\rm age} = 12.4$ Gyr \citep{simunovic_2023, choi_2025, cordoni_2025}. The track of the \gls{SMC}, going from the bottom-left to the top-right, as well as the field stars throughout, are clearly visible and have been successfully filtered out.}
    \label{fig:sample_selection}
\end{figure*}

One of the fields in Rubin \gls{DP1} was the 47 Tuc field, which is the focus of this work. The high stellar density in the cluster core of 47 Tuc saturated the CCD, which means that data is only available for the outer regions of the cluster ($\sim18-36$ arcmin, or $\sim5.7-11.4$ half-light radii) \citep{choi_2025, cordoni_2025}, as shown in the left panel of Figure \ref{fig:sample_selection}. There are a total of 72 exposures available in the \gls{DP1} data. However, no exposures were taken in the $z$ band for this field in Rubin \gls{DP1}. The $u$-band exposures also did not meet quality criteria and were excluded from the public \gls{DP1} release. Finally, the $y$-band only received 5 exposures, which means that many targets in the field do not have robust $y$-band photometry \citep{choi_2025}. Thus, we decided to work only with data from the $gri$ bands.

We accessed the Rubin \gls{DP1} \texttt{Object} table \citep{rubin_dp1_object_table} on the \gls{RSP} \citep{rubin_science_platform_2017, rubin_science_platform_2024} to query all objects within 2 degrees of the center of 47 Tucanae. Our sample selection is similar to \citet{choi_2025} and \citet{cordoni_2025}. To summarize, we explicitly excluded sources with \texttt{refExtendedness = 1}, which describes the extendedness parameter in the reference band, in order to remove galaxies and other non-stellar objects \citep{bosch_2018}. We also removed objects with \texttt{iPSF = True}, which denotes failure in the \gls{PSF} model moments, in any of the $gri$ bands. To avoid incorrect photometric estimates due to blending, we removed objects where the $\texttt{blendedness} > 0.05$ in any of the bands. We also excluded objects that are within 18 arcmin from the center to avoid crowding. The maximum distance of a cluster member to the cluster center in our sample is $\sim36$ arcmin. Assuming a half-light radius of $R_h = 3.17'$ \citep{harris_1996}, this coverage corresponds to a range of $5.7-11.4\;R_h$. This resulted in a parent sample with 10,136 objects, which is shown in grey in Figure \ref{fig:sample_selection}.

These thresholds guaranteed stellar sources with reliable photometry in the $gri$ bands in the 47 Tuc field. However, the exposures also included stars from the \gls{SMC} and overlapping field stars. To filter these out of our sample, we cross-matched our catalog with Gaia \gls{DR3} \citep{gaia_2016, gaia_dr3_2023} using the \texttt{coordinate\_matching} package\footnote{\url{https://github.com/tobiasgeron/coordinate_matching}} to ensure cluster membership using a maximum separation of \todo{1} arcsec and kept objects where the membership probability was greater than 0.8 \citep{vasiliev_2021}. We also created a PARSEC isochrone \citep{bressan_2012, chen_2015} using the web interface\footnote{\url{https://stev.oapd.inaf.it/cgi-bin/cmd}} assuming $d = 4.66$ kpc, $[M/H] = -0.5$, $E(B-V) = 0.025$, and $t_{\rm age} = 12.4$ Gyr \citep{simunovic_2023, choi_2025, cordoni_2025}\footnote{We use these values to stay consistent with \citet{choi_2025} and \citet{cordoni_2025}, two other studies that looked at 47 Tuc in Rubin \gls{DP1}. However, we note that the literature reports ages for 47 Tuc ranging between $\sim11 - 13$ Gyr \citep{thompson_2010,brogaard_2017, thompson_2020, simunovic_2023}, which would slightly alter the shape of the isochrone. However, as shown in Figure \ref{fig:sample_selection}, the isochrone closely traces the data, which suggests that the values used in this work are reasonable.}. We excluded stars from our sample if they have color differences larger than $\pm0.4$ from this isochrone. Finally, we exclude stars with $i < 17.5$ to remove stars beyond the main sequence turnoff point. We are left with a sample of \todo{1,499} stars in 47 Tucanae after applying these steps, visualized in the right panel of Figure \ref{fig:sample_selection} on an $i$ vs. $g-i$ \gls{CMD}.

\subsection{Stellar Model Templates}
\label{sec:template_library}

We use the catalog of stellar atmosphere models developed by \citet{castelli_2003}\footnote{\url{https://www.stsci.edu/hst/instrumentation/reference-data-for-calibration-and-tools/astronomical-catalogs/castelli-and-kurucz-atlas}}, which contains $\sim$4,300 models of stellar atmospheres for a broad range of effective temperatures ($3,500 \textrm{ K}< T_{\rm eff}< 50,000\textrm{ K}$), metallicities ($-2.5 < [M/H] < 0.5$), and surface gravities ($0.0 < \log(g \;[\rm{cm\;s^{-2}}]) < 5.0$). The wavelength range of the model spectra goes from the ultraviolet (1000 \AA) to the infrared (10 $\mu$m), which includes the entire wavelength range of the Rubin \gls{LSST} $ugrizy$ bands. These models were computed with the \texttt{ATLAS9} code, which uses the \gls{ODF} to calculate line opacities. The flux of these templates is in $F_\lambda$ surface flux units, i.e., erg s$^{-1}$ cm$^{-2}$ \AA$^{-1}$. They are also distance- and radius-independent (i.e., the flux is stored as the flux for an object with a radius of 1 and a distance of 1). This means that, to convert to observed fluxes, we need to multiply the template fluxes by a factor of $(R/D)^2$, where $R$ is the stellar radius and $D$ the distance to the star, in the same units.

While we use the stellar atmosphere models developed by \citet{castelli_2003}, the methods and tools in this work can be easily adapted to work with any library of stellar atmosphere models or custom templates.

\section{Autoencoder Framework and Training}
\label{sec:ae}

We train an \gls{AE} on a simulated dataset to classify single stars and binary systems using photometry from Rubin \gls{DP1}. The model architecture and loss function are described in Sections \ref{sec:ae_architecture} and \ref{sec:loss_function}, respectively. The simulated dataset we use to train the model is discussed in Section \ref{sec:creating_training_set}, while we describe how the uncertainty is modeled in more detail in Section \ref{sec:modeling_uncertainty}. Finally, the training procedure is described in Section \ref{sec:training}.

\subsection{Autoencoder architecture}
\label{sec:ae_architecture}

At its core, an \gls{AE} is an unsupervised learning model with three main separate components: an encoder, a latent space, and a decoder. The encoder tries to compress the input features, $x$, into a latent space, $z$. Meanwhile, the decoder uses the latent space to create a reconstruction of the input features, $\hat{x}$. By training the encoder and decoder together, an \gls{AE} learns to extract a representation of the input features in the latent space that contains all information necessary to reconstruct them. Ideally, this latent space representation is physically meaningful and can be used for other tasks, such as classification or anomaly detection. Autoencoders have been used widely in the astronomical literature (e.g., see \citealt{portillo_2020, villar_2020, villar_2021, huijse_2025, laroche_2025}).

\begin{figure*}
    \centering
    \includegraphics[width=0.7\textwidth]{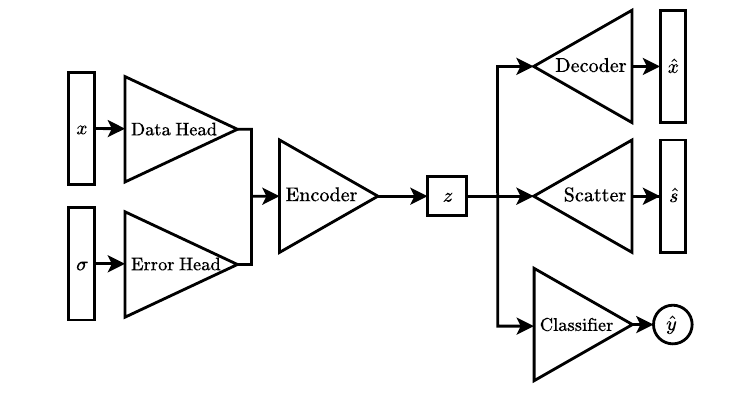}
    \caption{The model architecture of the \gls{AE} used in this work. It has the encoder and decoder components of a typical \gls{AE} that take the input features, $x$, and create the reconstructed features, $\hat{x}$. However, we also implement the scatter component described by \citet{laroche_2025} that outputs an additional scatter term, $\hat{s}$. We also added a separate input head that takes the uncertainty on the input features, $\sigma$, as a separate input before feeding it into the encoder. We also have a dedicated classifier component that makes predictions, $\hat{y}$, based on the latent space representation, $z$.}
    \label{fig:architecture}
\end{figure*}

In this work, we will use an \gls{AE} to classify targets as single stars or binary systems using photometric data from Rubin \gls{LSST} as input for the encoder. The complete \gls{AE} model architecture used in this work is shown in Figure \ref{fig:architecture}. The basic \gls{AE} components (encoder, decoder, and latent space) are clearly visible.

However, the uncertainties in real data are often heteroscedastic. Furthermore, the typical uncertainty for each band differs in Rubin \gls{LSST} (and many other astronomical surveys). We want our \gls{AE} to explicitly handle these complex uncertainties, which is done by adding the uncertainties of the features ($\sigma$) as separate inputs to the model. This is implemented by using two separate heads (called the data head and the error head in Figure \ref{fig:architecture}) before combining them into the encoder. 

We also implement the scatter component described by \citet{laroche_2025}, which is essentially a second decoder component, which is completely separate from the main decoder, that predicts model uncertainties. Importantly, the scatter component does not attempt to predict $\sigma$, i.e., the uncertainties on the input features, but instead captures the uncertainty on the model predictions for each target. However, as noted by \citet{laroche_2025}, interpreting exactly what kind of uncertainties the scatter component quantifies is hard. The scatter is essentially a metric that combines the intrinsic uncertainty of individual targets, systematics in the training data that are not represented by the input uncertainties, and outliers in the training data. \citet{laroche_2025} empirically found that including this component increased model performance, despite it being hard to interpret, which we have found to be true in this work as well.

We also add a trainable classifier component to the model architecture, similar to \cite{gomez_bombarelli_2016}. This classifier takes the latent space representation and makes a prediction ($\hat{y}$). This classifier is trained in tandem with the other components, which guarantees that the latent space is organized in such a way that is optimal for classification. 

In the implementation used in this work, the data and error heads are relatively simple components; they both contain only one layer. They each transform the input features and their uncertainties into a 12-dimensional vector. The main encoder component then takes this 24-dimensional vector (12 dimensions from each head) and feeds it through three layers, which compresses the data to 16, 8, and finally into the four-dimensional latent space.\footnote{We study the effect of the number of latent space dimensions on the performance of the model in Appendix \ref{app:size_latent_space}. We found that a four-dimensional latent space works well for the purposes of this work.} Every intermediate layer passes through a \gls{GELU} activation function. The decoder and scatter decoder are mirrored versions of the encoder. The scatter decoder contains a final additional sigmoid activation function to ensure positivity. The classifier component contains three layers of 16 neurons each, and a final fourth layer that compresses the data into one number, which is fed through a sigmoid activation function to ensure that it outputs a prediction with values between 0 (i.e., single star) and 1 (i.e., binary system).

In summary, the \gls{AE} used in this work has two types of inputs (the features, $x$, and their uncertainties, $\sigma$), three types of outputs (the reconstructed features, $\hat{x}$, the scatter, $\hat{s}$, and the classification, $\hat{y}$), six trainable components (the data head, error head, encoder, decoder, scatter, and classifier), and a total of \todo{1,779} trainable parameters. The effect of the size of the model and each component on the model performance is tested in more detail in Appendix \ref{app:model_size}. This model architecture is implemented with \texttt{PyTorch} \citep{pytorch_2019} and made publicly available in the \texttt{flexAE} package on GitHub.\footnote{\url{https://github.com/tobiasgeron/flexAE}} We want to highlight that, while we use this framework to identify \gls{MS}+\gls{MS} binary systems in Rubin \gls{DP1}, the \texttt{flexAE} code is designed so that users can adapt it to other science cases (e.g., see Appendix \ref{app:YSG} for a demonstration on using \texttt{flexAE} to find \gls{YSG} binaries) and that the aforementioned components and their sizes (i.e., number of layers and neurons in each layer) can be easily adjusted.

\subsection{The Loss Function}
\label{sec:loss_function}

The total loss of our model ($\mathcal{L}_{\rm tot}$) is basically the sum of two separate losses:

\begin{equation}
    \mathcal{L}_{\rm tot} = \mathcal{L}_{\rm rec} (x, \hat{x}, \sigma, \hat{s}) + \gamma \cdot \mathcal{L}_{\rm BCE} (y, \hat{y})\;,
    \label{eq:total_loss}
\end{equation}

where $\mathcal{L}_{\rm rec}$ is the reconstruction loss, $\mathcal{L}_{\rm BCE}$ the \gls{BCE} loss, and $\gamma$ is a proportionality coefficient. Similar to \citet{laroche_2025}, the reconstruction loss itself is a combination of two terms:

\begin{equation}
    \mathcal{L}_{\rm rec} (x, \hat{x}, \sigma, \hat{s}) = \chi^2(x, \hat{x}, \sigma, \hat{s}) + P(\sigma, \hat{s})\;,
    \label{eq:reconstruction_loss}
\end{equation}

where the first term is essentially a weighted \gls{MSE}: 

\begin{equation}
    \chi^2(x, \hat{x}, \sigma, \hat{s}) = \frac{1}{N} \sum_{i=1}^{N} \sum_{j=1}^{d} \frac{(x_{ij} - \hat{x}_{ij})^2}{\sigma_{ij}^2 + \hat{s}_{ij}^2} \;,
\end{equation}

where $N$ is the size of the batch, $d$ is the number of input dimensions, $x_{ij}$ is $j^{\rm th}$ input feature of the $i^{\rm th}$ target, $\hat{x}$ represents the reconstructed features, $\sigma$ the uncertainties on the input features, and $\hat{s}$ the scatter (see Section \ref{sec:ae_architecture}). The second term is equal to:

\begin{equation}
    P(\sigma, \hat{s}) = \frac{1}{N} \sum_{i=1}^{N} \sum_{j=1}^{d} \log (\sigma_{ij}^2 + \hat{s}_{ij}^2)\;.
\end{equation}

This is a penalty term added to the loss to prevent the model from simply predicting an infinite scatter to make the reconstruction loss zero. Note that the reconstruction can be negative. The first term of this equation will always be positive, as it is essentially a weighted mean squared error, but the second term will be negative if $\sigma^2 + \hat{s}^2 < 1$. We note that $\mathcal{L}_{\rm rec}$ in Equation \ref{eq:reconstruction_loss} can be derived directly from the negative log-likelihood of a Gaussian distribution with independent uncertainties.

The second term in Equation \ref{eq:total_loss} is the \gls{BCE} loss \citep{shannon_1948,good_1955, solla_1988} that ensures that the model is able to classify accurately: 

\begin{equation}
    \mathcal{L}_{\rm BCE}(y, \hat{y}) = -\frac{1}{N} \sum_{i=1}^{N} [y_i \cdot \log(\hat{y}_i) + (1 - y_i) \cdot \log(1 - \hat{y}_i)]\;,
    \label{eq:bce_loss}
\end{equation}

where $N$ is the size of the batch, $y_i$ is the true classification of the $i^{\rm th}$ target, and $\hat{y}_i$ is the predicted probability of the $i^{\rm th}$ target. The \gls{BCE} loss essentially corresponds to the negative log-likelihood of a Bernoulli distribution. Adding the \gls{BCE} loss to the total loss ensures that the \gls{AE} latent space is organized in such a way that is optimal for classification.

Note that the different losses inherently operate on different scales. For example, the \gls{BCE} loss for a classifier that guesses randomly is $\sim0.693$, while a perfect classifier will have a loss of 0. In contrast, the reconstruction loss can in principle go up to $+\infty$ if the reconstructed features are wildly incorrect. This is the reason behind the proportionality coefficient $\gamma$ in front of the \gls{BCE} loss in Equation \ref{eq:total_loss}, which allows users to finetune the weight of the \gls{BCE} loss relative to the reconstruction loss. We keep this at \todo{$\gamma = 20.0$} during the entire training process. The impact of the value of $\gamma$ on the model performance is investigated in more detail in Appendix \ref{app:finetune_gamma}.

\subsection{Creating a Simulated Training and Test Set}
\label{sec:creating_training_set}

It would be best to train and test the \gls{AE} on real data with accurate labels. However, such a sample is not always readily available, especially for new surveys such as Rubin \gls{LSST}. Instead, we use the stellar atmosphere models from \citet{castelli_2003} to generate a large sample of Rubin-like observations of \gls{MS} single stars and \gls{MS}+\gls{MS} binary systems in 47 Tuc.\footnote{While we use the templates from \citet{castelli_2003} and the Rubin \gls{LSST} photometric bands in this work, the \texttt{flexAE} framework is built so that users can easily specify their own templates and their photometric bands of choice.} A large advantage of this approach is that it easily allows us to generate an arbitrarily large training set for any combination of photometric bands.\footnote{If a dataset with real data with accurate labels is available for your use case, even if it is a small sample, you can always train with the large simulated template training set, and then finetune and/or test with the real data (for examples of finetuning and transfer learning in astronomy, see e.g., \citealt{walmsley_2022b, burhanudin_2023,gupta_2025}).}

\begin{figure}
	\includegraphics[width=\columnwidth]{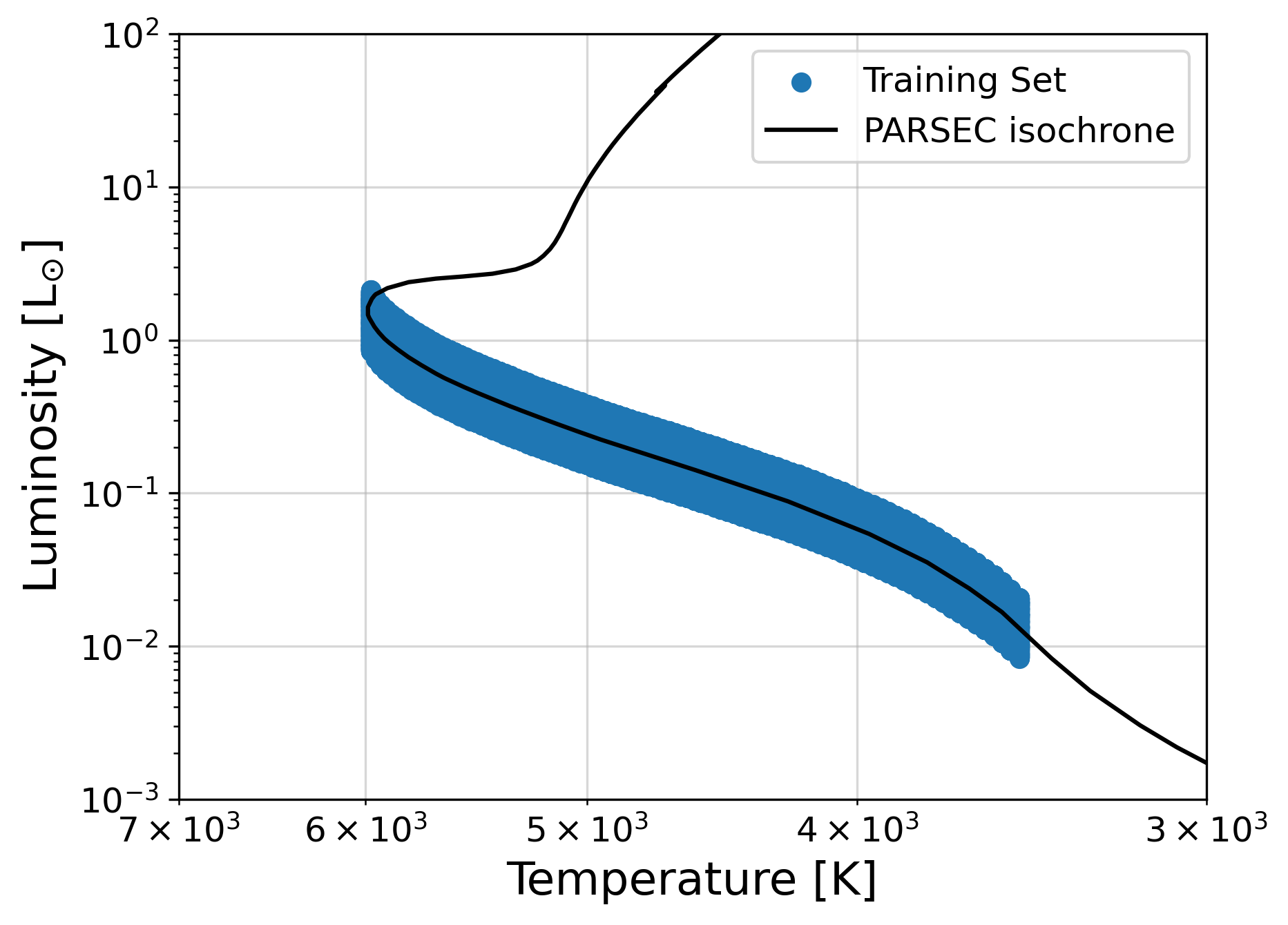}
    \caption{The positions of the simulated training set on the Hertzsprung–Russell diagram. We also show the PARSEC isochrone of 47 Tuc. The simulated training set encompasses the main sequence of the 47 Tuc isochrone.}
    \label{fig:HR_training_set}
\end{figure}

While creating the simulated training and test sets, we assume the same parameters that were used to create the PARSEC isochrone in Section \ref{sec:sample_selection}, i.e., $d = 4.66$ kpc, $[M/H] = -0.5$, $E(B-V) = 0.025$, and $t_{\rm age} = 12.4$ Gyr \citep{simunovic_2023, choi_2025, cordoni_2025}. We use stellar templates with temperatures between \todo{3,500} K and \todo{6,000} K. This temperature range encompasses the upper range of the \gls{MS} of the 47 Tuc isochrone, as shown in Figure \ref{fig:HR_training_set}. While the metallicity equals $-0.5$ for 47 Tuc, we include templates with $-2.5 < [M/H] < 0$ to increase the diversity of the training set. We also only choose templates with $4.0 < \log(g \;[\rm{cm\;s^{-2}}]) < 5.0$. The templates in \citet{castelli_2003} have step sizes of 250, 0.5, and 0.5 in temperature, metallicity, and surface gravity, respectively. We augment the sample by interpolating between templates down to a step size of 25, 0.1, and 0.1 for temperature, metallicity, and surface gravity. These thresholds result in a final set of $\sim$\todo{51,000} unique single star templates that we use to create our simulated training set. The spectra of these templates were convolved with the Rubin \gls{DP1} throughput curves of the $gri$ bands using the tools available in the \texttt{MultiStarFitting} package.

\begin{figure*}
	\includegraphics[width=\textwidth]{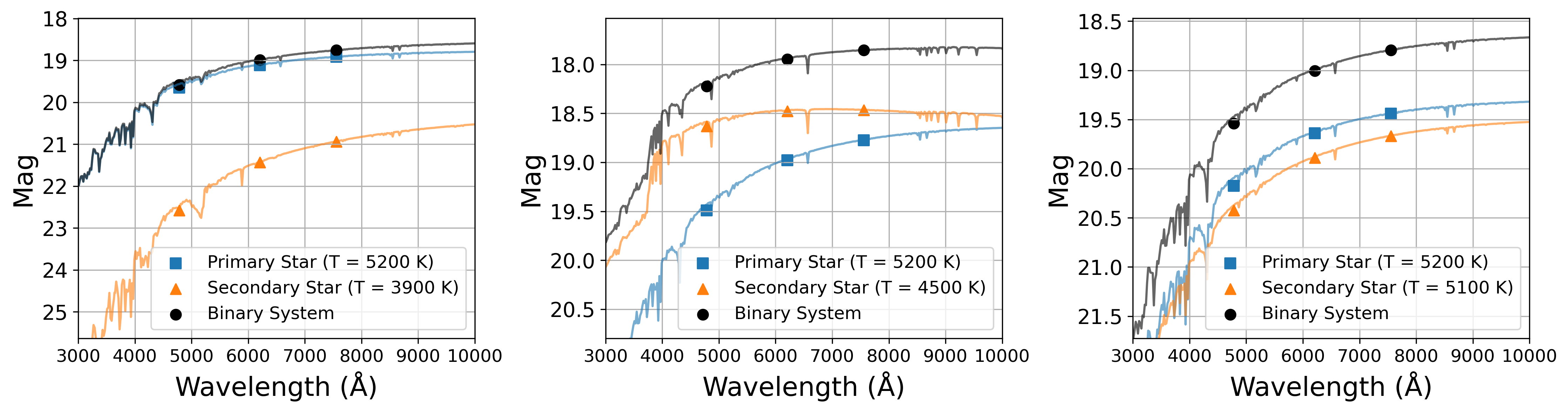}
    \caption{The $gri$ band magnitudes (black circles) and full PARSEC spectrum (black line) of a few examples of binary systems in our training set. The band magnitudes and full theoretical spectra of the primary star (blue squares) and secondary star (orange triangles) are also shown. The binary systems tend to be redder and brighter than their primary stars by themselves.}
    \label{fig:broadband_sed_examples}
\end{figure*}

To generate our simulated set of single stars, we sample from a \citet{kroupa_2001} \gls{IMF} and use the 47 Tucanae isochrone to convert stellar masses to effective temperatures. We then select a template from the template library with a temperature that is closest to the sampled temperature. We can then determine its position on the Hertzsprung–Russell diagram by using the 47 Tucanae isochrone to estimate the corresponding luminosity. To increase the diversity of the training set, we allow the luminosity to vary by $\pm0.2$ dex. The fluxes for each template were adjusted for the distance to 47 Tuc and for the radius of the star based on its position on the Hertzsprung–Russell diagram, as the templates are distance- and radius-independent (see Section \ref{sec:template_library}).\footnote{The mass of most (i.e., the 10$^{\textrm{th}}$ to 90$^{\textrm{th}}$ percentile) of the stars in our simulated set varies between $\sim0.12$ $M_{\odot}$ and $\sim1.26$ $M_{\odot}$. However, due to the range of surface gravities used ($4.0 < \log(g \;[\rm{cm\;s^{-2}}]) < 5.0$) and the additional variation in luminosity by $\pm0.2$ dex, there are a few outliers, ranging between $\sim0.02$ $M_{\odot}$ and $\sim6.4$ $M_{\odot}$.} 

For binary systems, we first select a template that corresponds to the primary star in the same manner as described above. However, we then randomly choose the template for the secondary star by uniformly sampling all templates while additionally enforcing that the secondary star has a lower temperature than the primary to approximate a flat mass ratio distribution (defined in this work as $q = M_2/M_1$, where $M_1$ is the stellar mass of the primary star and $M_2$ the stellar mass of the secondary star). We then simply add the fluxes of both components together.

Finally, the template fluxes are adjusted for extinction assuming a \citet{odonnell_1994} extinction model with $E(B-V) = 0.025$ and $R_V = 3.1$ (and thus $A_V = 0.078$). To allow for additional variation in the training set, we allow $A_V$ to vary by $\pm0.1$. Since we are using Gaia \gls{DR3} to determine cluster membership, we are limited to targets brighter than $i \sim 20.5$ mag (see Section \ref{sec:sample_selection} and Figure \ref{fig:sample_selection}). Thus, we only include targets in the simulated set if $i < 21$ mag. This is repeated until we obtain a sample of \todo{50,000} targets, split equally between single stars and binary systems. This simulated dataset is randomly divided into training and test sets using an 80/20 split, so that the training set contains \todo{40,000} targets and the test set contains \todo{10,000} targets. The effect of the training set size on model performance is studied in more detail in Appendix \ref{app:training_set_size}.

The $gri$ magnitudes and the full theoretical spectra of a few examples of binary systems generated in our simulated training set are shown in Figure \ref{fig:broadband_sed_examples}. As expected, the binary systems tend to be redder and brighter than their primary stars alone. This is what we anticipate the model will learn during the training process to distinguish between single stars and binary systems.

\subsection{Modeling the Uncertainty}
\label{sec:modeling_uncertainty}

It is essential to model the relationship between the template band fluxes and their uncertainties correctly, as otherwise the \gls{AE} will not be able to generalize well from the simulated data to the real data. In this work, we follow the error model presented in Section 3.2.1 of \citet{ivezic_2019}. They model the expected photometric error in magnitudes as:

\begin{equation}
    \sigma^2_{\rm tot} = \sigma^2_{\rm sys} + \sigma^2_{\rm rand}\;,
    \label{eq:total_error}
\end{equation}

where $\sigma^2_{\rm sys}$ is the systematic error and $\sigma^2_{\rm rand}$ is the random photometric error. The latter is described as: 

\begin{equation}
    \sigma^2_{\rm rand} = (0.04 - \gamma) x + \gamma x^2\;,
    \label{eq:random_error}
\end{equation}

\begin{deluxetable*}{c c c c}
  \tablecaption{The parameters that define the uncertainty for any given input magnitude using the error model presented by \citet{ivezic_2019} and shown in Equations \ref{eq:total_error} and \ref{eq:random_error} for each band in the Rubin \gls{DP1} 47 Tuc field. We estimated $\sigma_{\rm sys}$ by calculating the median error for all bright ($m<17$) sources in the Rubin \gls{DP1} 47 Tuc field separately for each band. We obtained $m_5$ from \cite{choi_2025} and $\gamma$ from \citet{ivezic_2019}.}
  \label{tab:error_model_parameters}
  \tablehead{
    \colhead{\textbf{Model Parameter}} & \multicolumn{3}{c}{\textbf{Band}} \\
    \cmidrule(lr){2-4}
    \colhead{} & \colhead{$g$} & \colhead{$r$} & \colhead{$i$}
    }
  \decimalcolnumbers
  \startdata
    $\sigma_{\rm sys}$ & 0.0007 & 0.0005 & 0.0005 \\
    $m_5$ & 24.14 & 24.14 & 23.77 \\
    $\gamma$ & 0.039 & 0.039 & 0.039 \\
  \enddata
\end{deluxetable*}

\begin{figure*}
	\includegraphics[width=\textwidth]{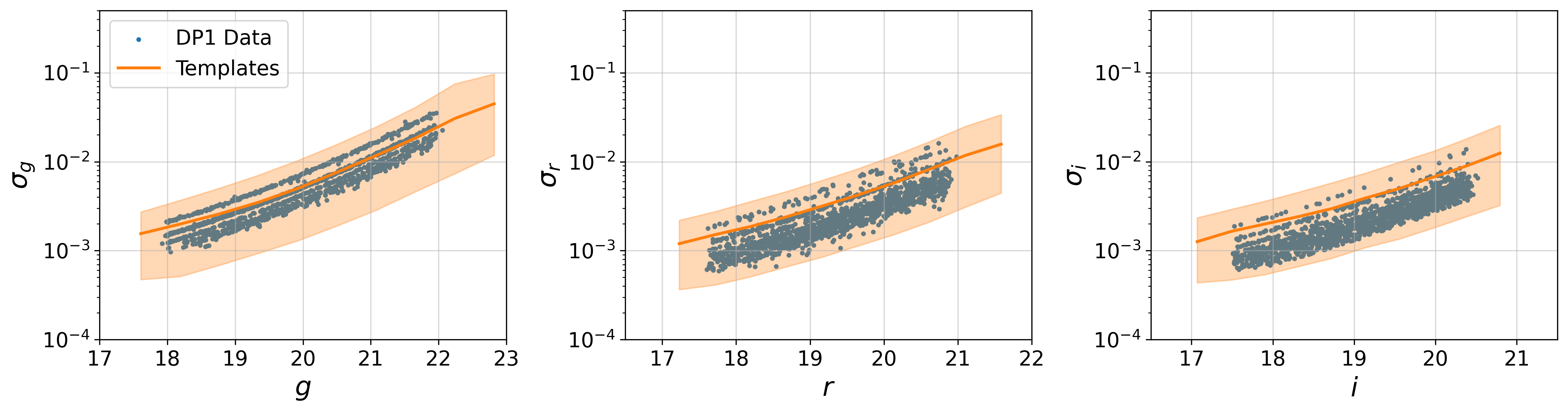}
    \caption{The uncertainties for each band plotted against the corresponding band magnitude for each of the $gri$ bands available for 47 Tuc in the Rubin \gls{DP1} data. The 47 Tuc data is shown in blue, while the data for the simulated templates used in this work is shown in orange.}
    \label{fig:comparison_errors}
\end{figure*}

where $\gamma$ is dependent on factors such as the sky brightness and readout noise. We also define $x = 10^{0.4 (m - m_5)}$, where $m$ is the expected magnitude and $m_5$ the $5\sigma$ point source depth in a given band. \citet{ivezic_2019} note that $\gamma$ equals 0.038 for the $u$ band, while it equals 0.039 for all other bands in Rubin \gls{LSST} (i.e., the $grizy$ bands). We use the $5\sigma$ point source depths found for Rubin \gls{DP1} in the 47 Tuc field as quoted in Table 1 of \citet{choi_2025}. Additionally, \citet{ivezic_2019} note that the calibration systems and procedures are designed to keep $\sigma^2_{\rm sys} < 0.005$ mag. However, to make sure we correctly model the uncertainty, we take a closer look at sources in the 47 Tuc field of Rubin \gls{DP1} with magnitudes brighter than 17 mag. As $\sigma^2_{\rm rand}$ decreases as sources get brighter (see Equation \ref{eq:random_error}), $\sigma^2_{\rm sys}$ will start to dominate for these bright sources. We assume that the median error of all sources brighter than \todo{17} mag in each band approximates $\sigma_{\rm sys}$. All the parameters we assumed in this error model are listed in Table \ref{tab:error_model_parameters}. Finally, to account for any unforeseen systematic biases in the uncertainties that will appear in real data and to provide additional variation in the training data, we multiply the final uncertainty by a random number between \todo{0.25 and 1.4}.


The uncertainty in each band is plotted against the magnitude in that band in Figure \ref{fig:comparison_errors} for the three bands used in this work ($gri$). The uncertainties for the real \gls{DP1} data are shown in blue, while the range of values for the simulated templates used in this work is shown in orange. As expected from Equations \ref{eq:total_error} and \ref{eq:random_error}, the uncertainty increases as the target becomes fainter. Crucially, the simulated templates encapsulate the uncertainties of the real Rubin \gls{DP1} data for all magnitudes, which ensures that the uncertainties of the simulated test set are comparable to the uncertainties in the real data. 

We want to stress that the distribution of the simulated uncertainties should match the observed uncertainty distribution. If this is not the case, then the observations will essentially be out-of-distribution for the trained model. This will influence the reconstruction and eventually the classification, as the error head, by design, shapes the latent space and two otherwise identical targets with different uncertainties will be placed at different locations in the latent space.

We also want to highlight that the uncertainties between the bands in the real data are correlated. If users find that the \gls{AE} classifier is struggling to generalize from the simulated test set to real data for their science case (e.g., in terms of the reconstruction and/or distance to nearest neighbor, see Section \ref{sec:latent_space_outliers}), then correlations in the uncertainties are a likely culprit. If this happens, we strongly encourage users to plot the uncertainties of all bands against each other for the simulated test set and real data, to see if the simulated test set encapsulates all possible combinations of model uncertainties and their correlations. This is especially the case if the generalization from the simulated test set to real data works significantly better if the separate error head is disabled.

\subsection{Training the Autoencoder}
\label{sec:training}

\begin{figure*}
	\includegraphics[width=\textwidth]{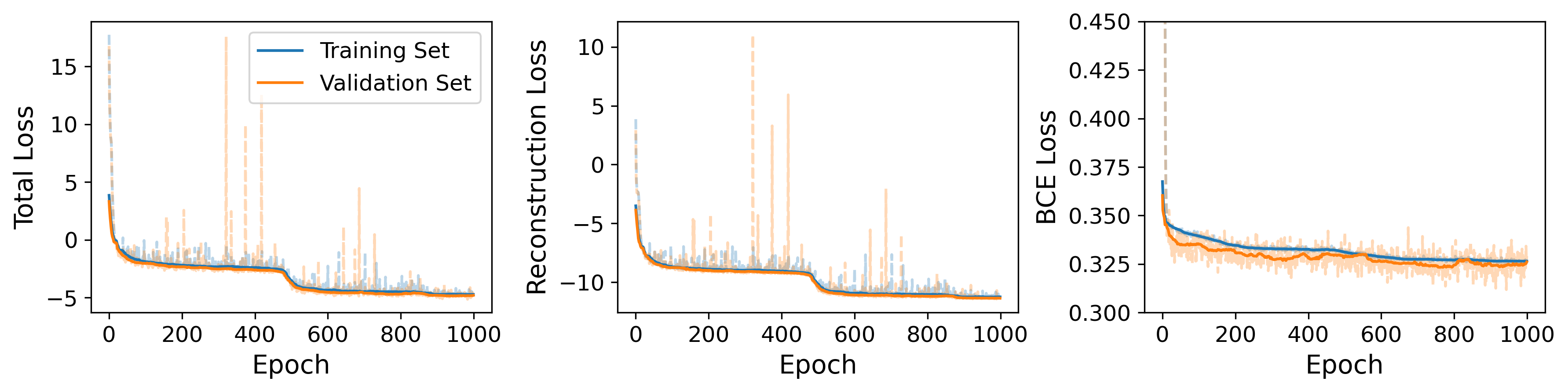}
    \caption{The left panel shows the total loss curve for the training (blue) and validation (orange) sets. The transparent dashed lines represent the raw loss curves, whereas the solid lines show the rolling median values with a window size of \todo{40} epochs. The reconstruction loss curve for the training and validation sets is shown in the middle panel, while the right panel shows the \gls{BCE} loss curve. The loss curve for all three losses has flattened at a minimum after training for \todo{1,000} epochs.}
    \label{fig:loss_curve}
\end{figure*}

As noted in Section \ref{sec:sample_selection}, we use the $gri$ bands from Rubin \gls{DP1}. We combine these three bands into the following two colors that capture the broadband \gls{SED} of each target, which are used as input features for the \gls{AE}: $g-r$ and $r-i$. We also add the $r$-band magnitude as a third input feature to anchor the vertical scaling of the broadband \gls{SED} of each target. Thus, we have 6 total input features: three data features ($r$, $g-r$, and $r-i$) and their associated uncertainties ($\sigma_{r}$, $\sigma_{g-r}$, and $\sigma_{r-i}$). All of the input data features are standardized to zero mean and unit variance by subtracting the mean and dividing by the standard deviation. To keep the relative ratio of the data features to the corresponding uncertainty features intact, the latter are not standardized in the same way. Instead, the uncertainty features are divided by the standard deviation of their corresponding data feature.

The \gls{AE} was trained for \todo{1,000} epochs. The total loss, reconstruction loss, and \gls{BCE} loss over each epoch are shown in Figure \ref{fig:loss_curve}. The loss curve for all three losses has flattened after training for \todo{1,000} epochs. We use an \texttt{AdamW} optimizer and a \texttt{ReduceLROnPlateau} learning-rate scheduler with a starting learning rate of \todo{1e-3}. The batch size equals \todo{640} and we have a validation split of \todo{0.1}. The training took $\sim$\todo{30} minutes on the Rubin Science Platform using the ``large'' environment setting (i.e., 4.0 CPUs and 32Gi RAM). The \gls{AE} model parameters from the epoch with the lowest validation loss are used for the rest of this work.

\section{Results}
\label{sec:results}

\subsection{Model Performance on Simulated Templates}
\label{sec:results_model_performance_on_sims}

\begin{figure*}
\centering
\gridline{\fig{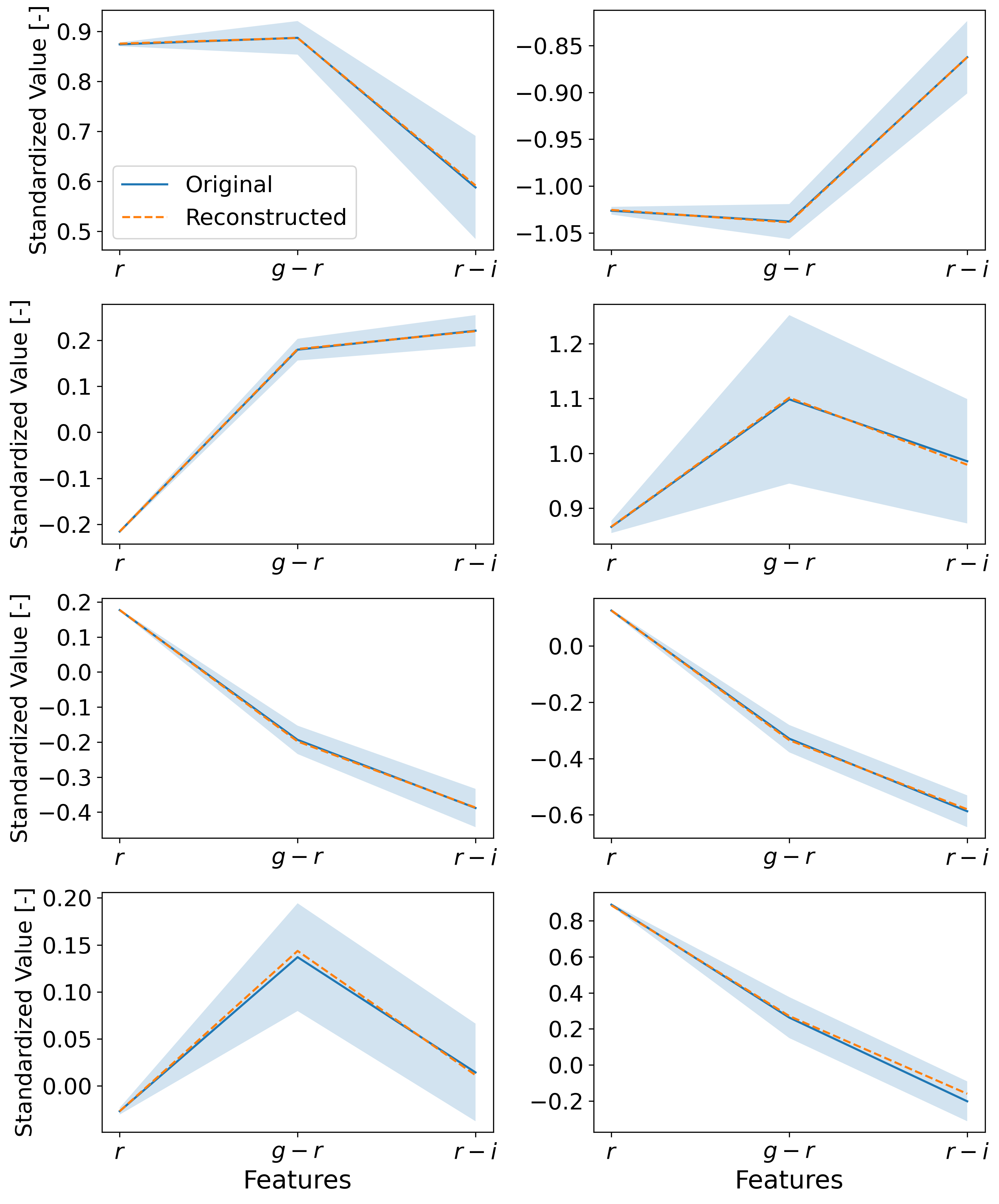}{0.48\textwidth}{}
          \fig{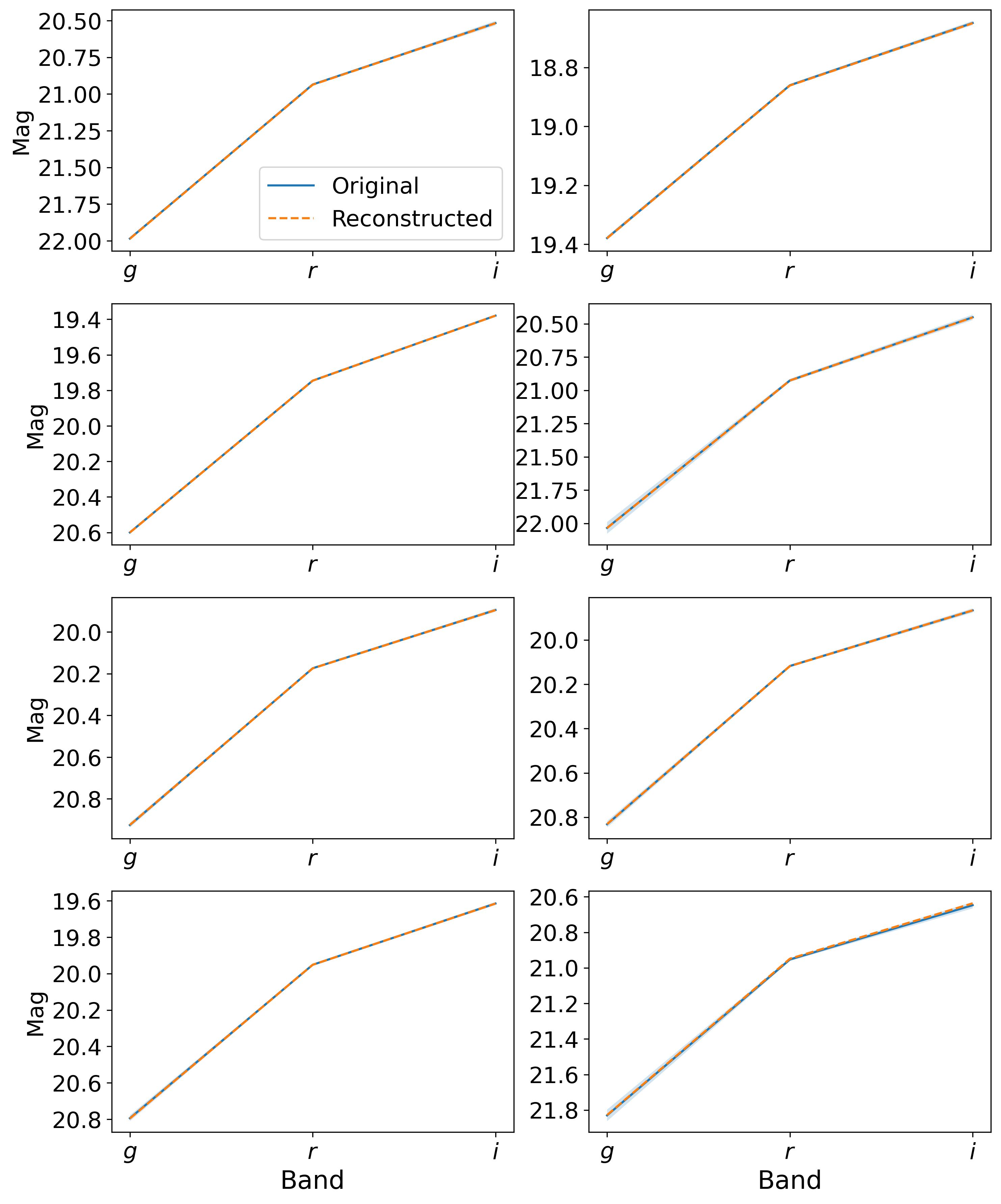}{0.48\textwidth}{}}
\caption{The left multi-panel plot shows the values of the standardized input features (blue) as well as the reconstructed \gls{AE} output features (orange) for a random subset of targets from the training set. The shaded blue regions represent the uncertainties on the features. The right multi-panel plot shows the unscaled $gri$ bands and the unscaled reconstructed band magnitudes for the same targets, which are calculated by combining and unscaling the reconstructed features. The \gls{AE} reconstruction is similar to the input.}
\label{fig:features_reconstructed}
\end{figure*}

As noted above, we train the model on six input features: the $g-r$, $r-i$ colors, the $r$-band magnitude, and their associated uncertainties. The standardized input features for a random subset of examples in the training set are shown in blue in the left panel of Figure \ref{fig:features_reconstructed}, while the reconstructed standardized features are shown in orange. Based on visual inspection of these examples, the model seems to be able to reconstruct the features well. This suggests that the model learned to represent the required information for reconstruction in the four-dimensional latent space. The right panel of Figure \ref{fig:features_reconstructed} shows the unscaled input $gri$ bands and the unscaled reconstructed $gri$ bands, which highlights again that the \gls{AE} model is able to reconstruct the targets well.

\begin{figure}
	\includegraphics[width=\columnwidth]{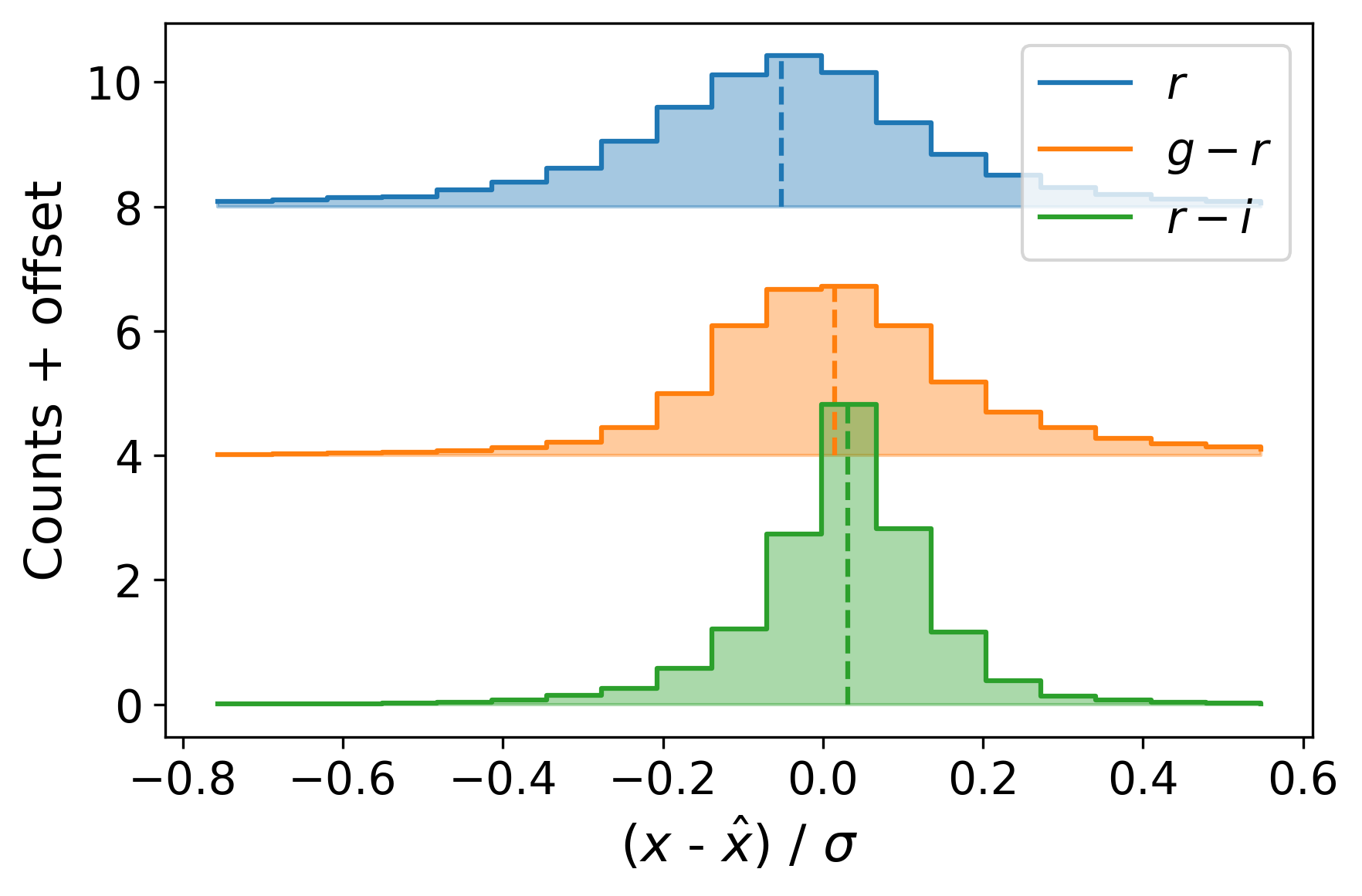}
    \caption{The number of standard deviations that the reconstructed features are removed from the input features (i.e., $(x - \hat{x}) / \sigma$) for each target in the training set. This is shown separately for each feature: the $r$-band magnitude (top; blue), the $g-r$ color (middle; orange), and $r-i$ color (bottom; green). The vertical dashed lines denote the median for each distribution. Note that the distributions are offset vertically to aid visualization.}
    \label{fig:n_std}
\end{figure}

The reconstructed features are always within the uncertainties of the input features for the few examples shown in Figure \ref{fig:features_reconstructed}. The quality of the reconstruction of the entire training set is quantified in greater detail in Figure \ref{fig:n_std}, where we calculate the number of standard deviations the reconstructed features are removed from the input features for each target in the training set. The distribution is centered around 0 for all three features, and rarely extends beyond 1. The reconstructed features are removed from the input features by more than one standard deviation for only \todo{$\sim0.8$}\% of the cases. In fact, \todo{$\sim89$}\% of the time, the reconstruction was within $0.3$ standard deviations of the input features. Thus, we conclude that \gls{AE} behaves as expected and that the model is able to reconstruct our targets well.

\begin{figure*}
	\includegraphics[width=\textwidth]{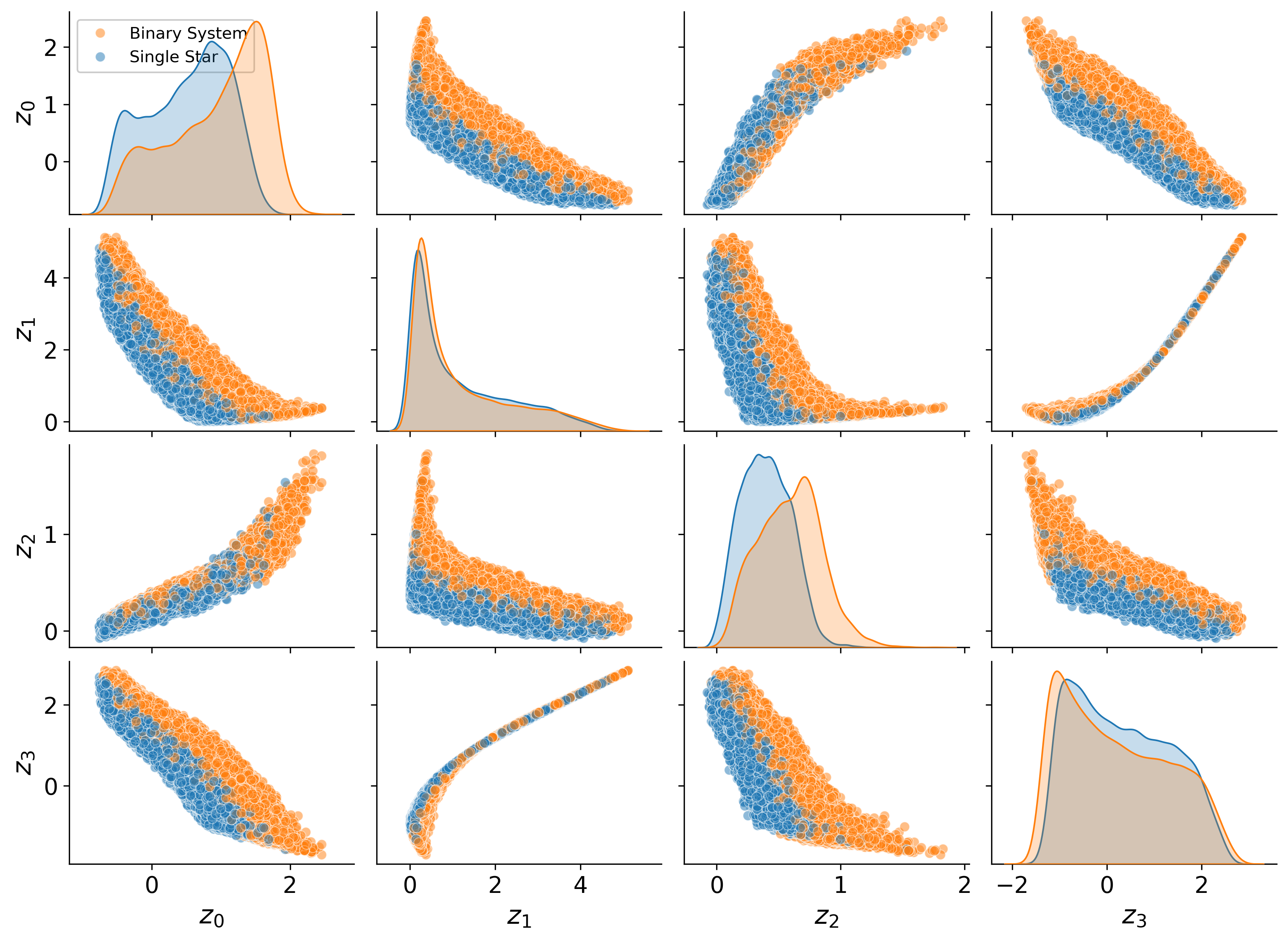}
    \caption{The location of the training sample in the \todo{four}-dimensional latent space of the model. Each point is colored by whether it is a single star (blue) or a binary system (orange). The two classes are separated well in this latent space.}
    \label{fig:latent_space}
\end{figure*}

\begin{figure}
	\includegraphics[width=\columnwidth]{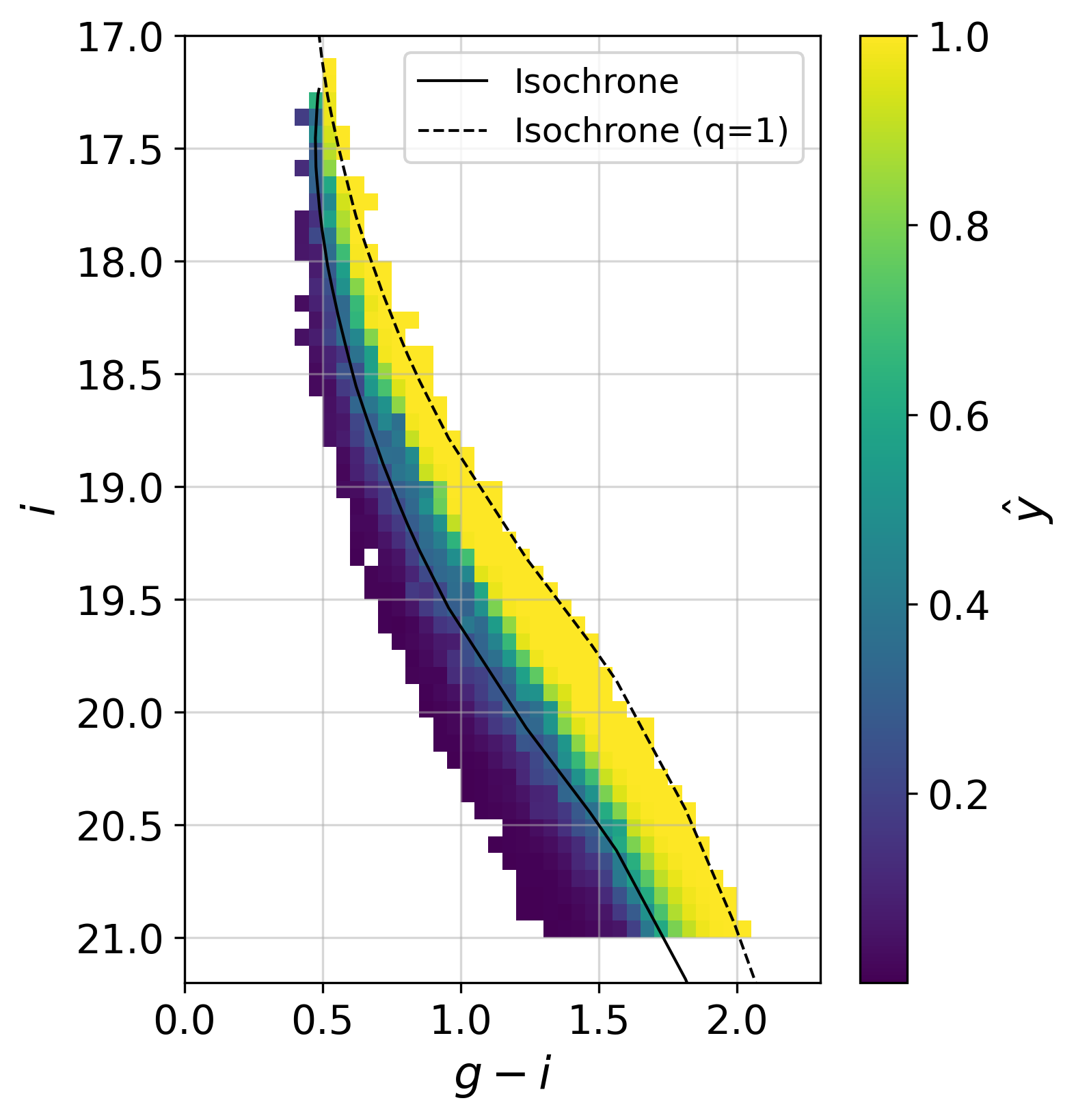}
    \caption{The median value of $\hat{y}$ for different locations on the $i$ vs. $g-i$ \gls{CMD} for the simulated test set. The full line represents the main sequence PARSEC isochrone, while the dashed line represents the $q=1$ isochrone.}
    \label{fig:cmd_model_predictions_locations}
\end{figure}

We mostly care about the ability of the model to use its latent space to classify targets correctly as single stars or binary systems. The location of the training sample in the \todo{four}-dimensional latent space of the model is shown in Figure \ref{fig:latent_space}, where the single stars are colored in blue and the binary systems are colored in orange. The latent space separates the single stars from the binary systems well. Figure \ref{fig:cmd_model_predictions_locations} shows the median value of $\hat{y}$ for the simulated test set on the $i$ vs. $g-i$ \gls{CMD}. As expected, the value of $\hat{y}$ increases from 0.0 to 1.0 as we move from the main sequence isochrone to the $q=1$ isochrone.

\begin{figure}
	\includegraphics[width=\columnwidth]{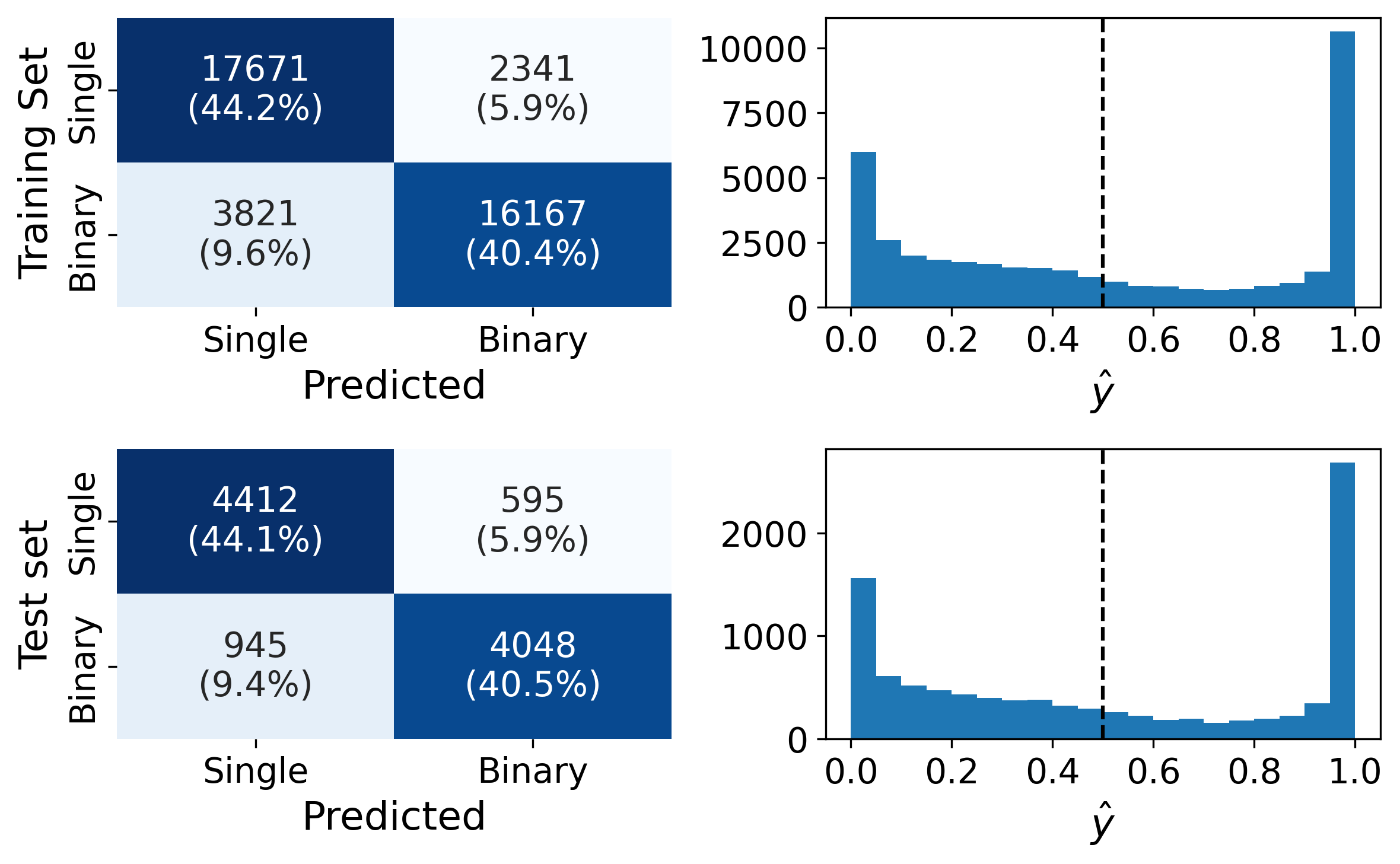}
    \caption{The confusion matrix of the real and predicted classes in the training and test samples (top-left and bottom-left panels, respectively). Most targets lie on the diagonal in both confusion matrices. The classification accuracy is \todo{0.85} for both samples, which suggests that the model did not overfit. The full distribution of $\hat{y}$ for the training and test samples is shown in the top-right and bottom-right panels, respectively.}
    \label{fig:classification}
\end{figure}

The confusion matrices for the training and test sets are shown in the left column of Figure \ref{fig:classification}. Most targets lie on the diagonal for both the training and test sets. The classification accuracy is \todo{0.85} for both samples, which suggests that the model did not overfit. These confusion matrices were created using a $\hat{y}$ threshold of 0.5. The full distribution of $\hat{y}$ for the training and test samples is shown in the top-right and bottom-right panels, respectively. These show that the distribution of $\hat{y}$ for both samples is very bimodal. This suggests that the model is often confident that the target is either a single star (i.e., low $\hat{y}$) or a binary system (i.e., high $\hat{y}$).

\begin{figure}
	\includegraphics[width=\columnwidth]{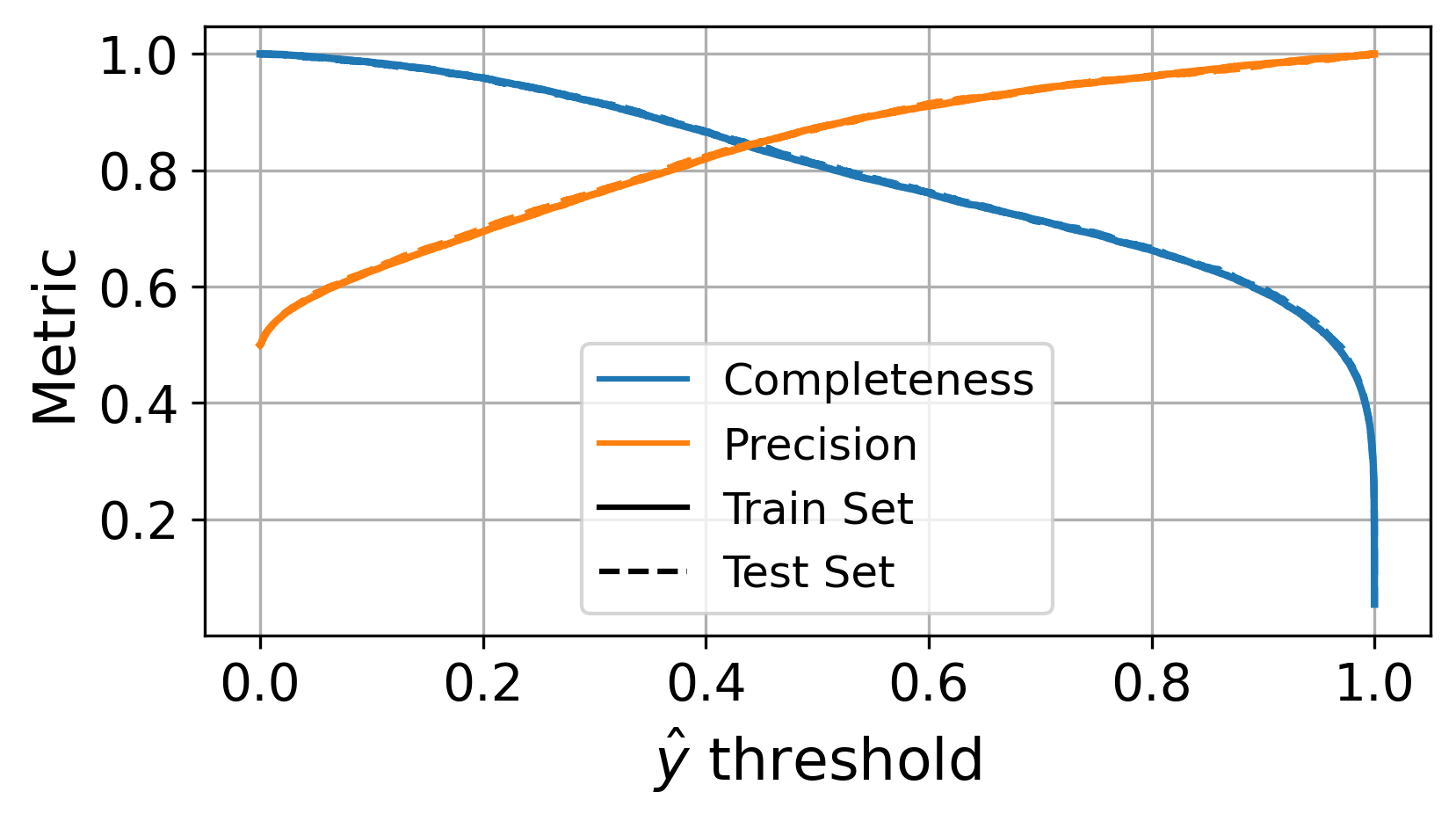}
    \caption{The precision (orange) and completeness (blue) of the binary class for the training (solid line) and test (dashed line) samples.}
    \label{fig:precision_completeness}
\end{figure}

As shown in Figure \ref{fig:precision_completeness}, it is possible to increase precision at the cost of completeness by only including targets with confident classifications. For example, the precision and completeness for the binary class in the test set equals \todo{0.87} and \todo{0.81}, respectively, using a threshold of $\hat{y} > 0.5$. This changes to \todo{0.98} and \todo{0.60}, respectively, using a stricter threshold of $\hat{y} > 0.9$ instead.

\begin{figure}
	\includegraphics[width=\columnwidth]{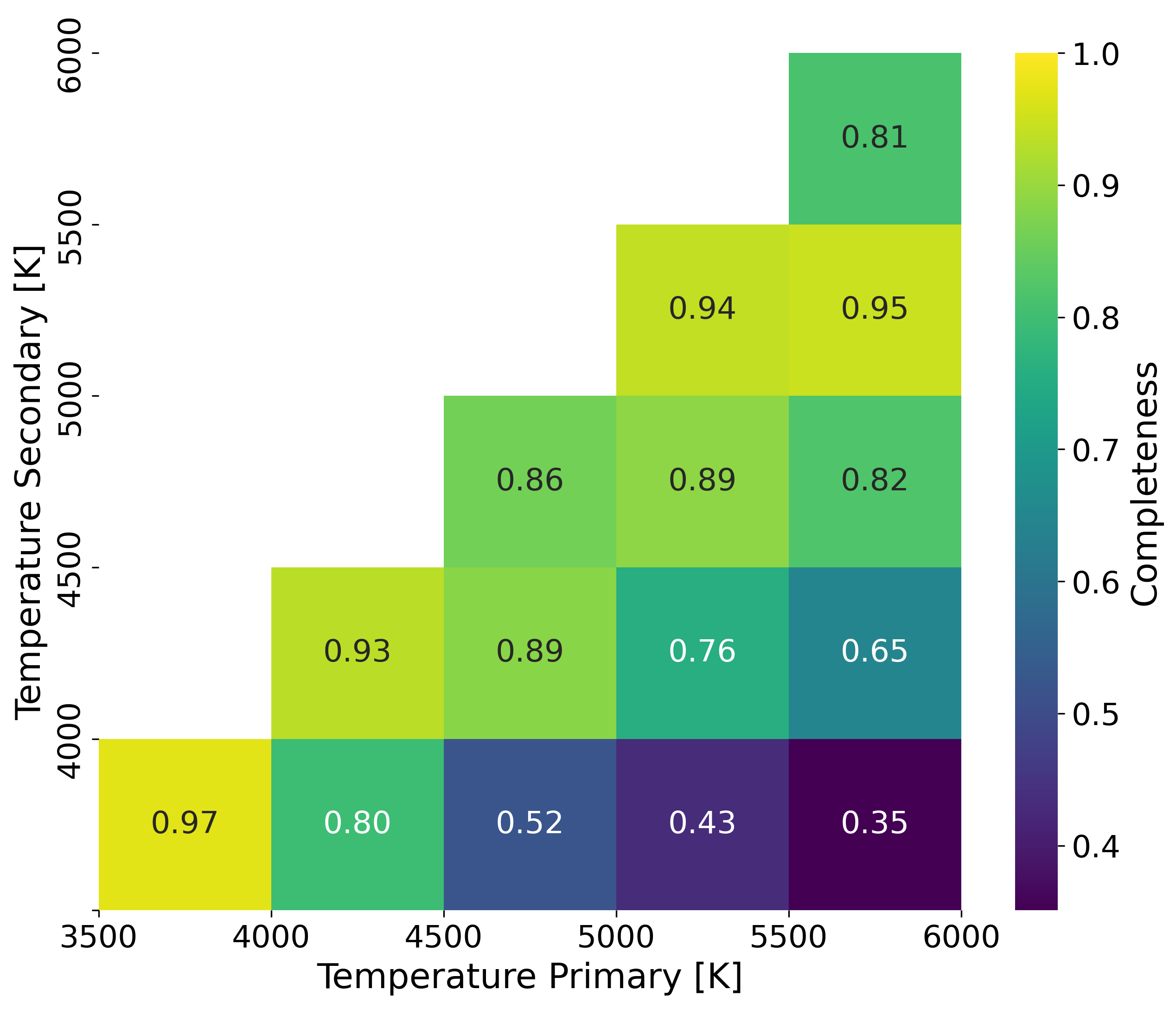}
    \caption{The completeness of the binary classification (assuming $\hat{y} > 0.5$) in the test set as a function of the temperature of the primary star and the secondary star. As we are considering \gls{MS}+\gls{MS} binaries, this directly corresponds to luminosity- and mass-ratios.}
    \label{fig:completeness_2d}
\end{figure}

The completeness of the binary classification (using $\hat{y} > 0.5$ to select binary systems) in the test set as a function of the temperature of the primary star and its companion is shown in Figure \ref{fig:completeness_2d}. As expected, the completeness depends on both temperatures. It is generally harder to find a binary with a 3,500 K companion star, compared to a binary with a 5,000 K companion star. For any given primary star temperature, the completeness tends to go up if the temperature of the secondary star increases. Conversely, it is easier to spot a fainter companion if the primary star has a lower temperature; the completeness is much higher for a binary system with a 4,500 K primary star and 3,500 K companion, than a binary system with a 6,000 K primary star and 3,500 K companion. We emphasize that this is not inherently a temperature-dependent phenomenon, but instead relates to the relative flux contribution of both stars to the total flux, which happens to be correlated with temperature for the \gls{MS}+\gls{MS} binaries considered in this work. In other words, the brighter primary star will completely outshine the fainter star, which makes it harder to identify. Binary systems where both stars have similar temperatures (and, for \gls{MS} stars, high mass ratios of $q \gtrsim 0.7$) are easier to find than binaries where the primary star is much hotter and brighter than the secondary star (i.e., low mass ratios of $q \lesssim0.7$).

We want to reiterate that the values of $\hat{y}$, precision, and completeness presented in Figures \ref{fig:classification}, \ref{fig:precision_completeness}, and \ref{fig:completeness_2d} are derived from the simulated set based on stellar atmosphere models. They serve as a useful proxy for model performance. However, these values are not necessarily representative of what the precision and completeness would look like for real data. Consequently, it is not clear that the completeness values from Figure \ref{fig:completeness_2d} can be used to, e.g., correct binary fractions obtained from the Rubin \gls{DP1} data (see Section \ref{sec:dp1_binary_47tuc}). This is partially because simulated templates are not perfect and will deviate from the real data in subtle ways. Additionally, differences in the distribution of target parameters (e.g., in the mass of the primary star and mass ratio) between the training set and the real data will also affect model performance metrics. Finally, as noted in Section \ref{sec:creating_training_set}, we also included targets in the simulated set with slightly different metallicities, surface gravities, and $A_V$ values than those found in the real 47 Tucanae population. This was done to add variability in the training set, which helps the \gls{AE} to learn the underlying physical relationships more robustly, but this can cause differences in the model performance metrics between the simulated and real data as well.

\subsection{Detection of Outliers in Latent Space}
\label{sec:latent_space_outliers}

\begin{figure}
	\includegraphics[width=\columnwidth]{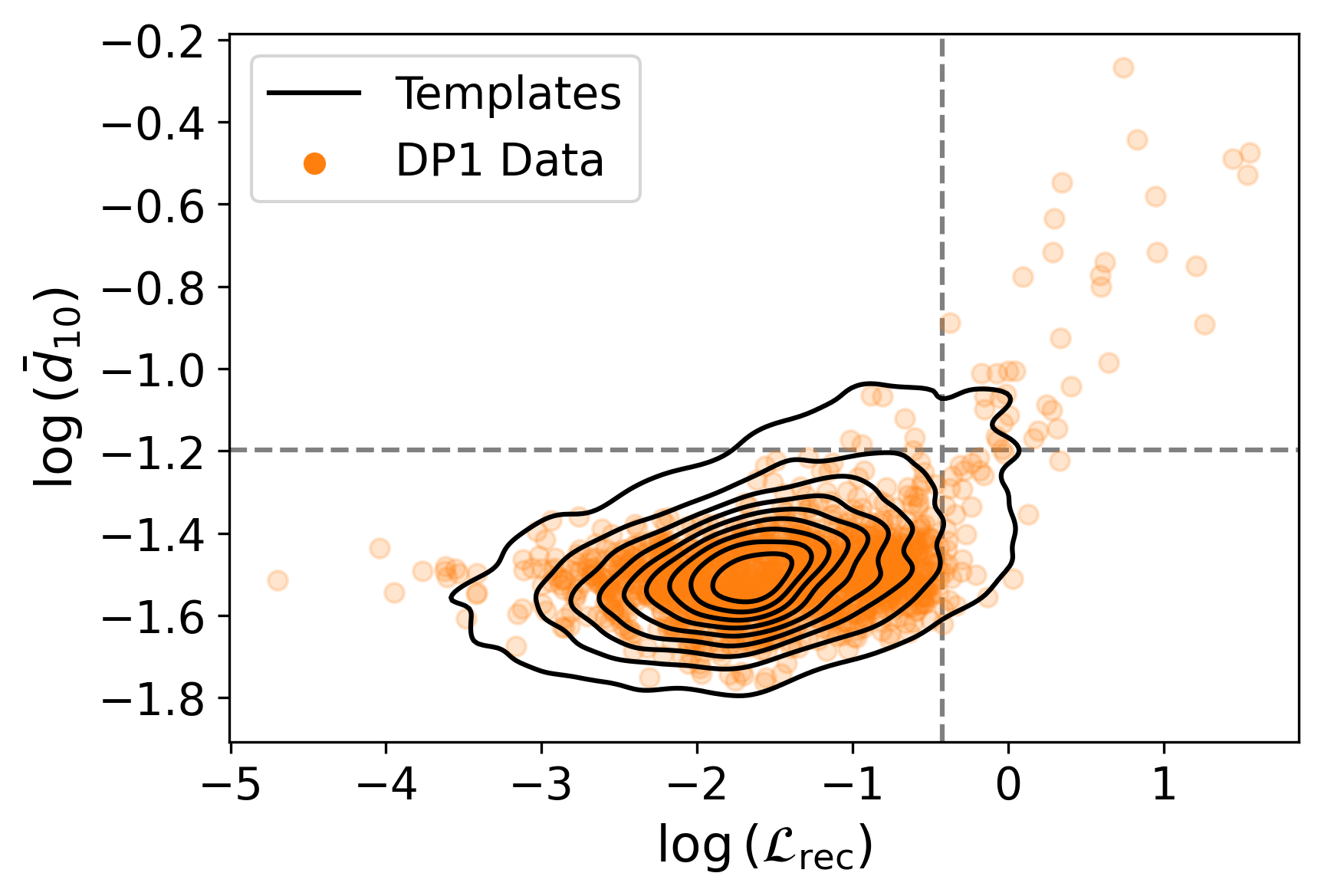}
    \caption{The logarithm of the reconstruction loss, $\log{(\mathcal{L}_{\rm rec})}$, against the logarithm of the mean distance to the 10 nearest neighbors in the latent space, $\log{(\bar{d}_{10})}$, of each target in the simulated template training set (black contours) and the real Rubin \gls{DP1} set (orange markers). The horizontal and vertical grey dashed lines are placed at the 97$^{\rm th}$ percentile of the simulated template training set. The real and simulated sets have similar values for $\log{(\mathcal{L}_{\rm rec})}$ and $\log{(\bar{d}_{10})}$, which suggests that the simulated templates are representative of the real data.}
    \label{fig:latent_outliers}
\end{figure}

The \gls{AE} has now been trained on the simulated training set. However, before applying the model to the real Rubin \gls{DP1} data, we need to ensure that it generalizes well from the simulated training set to the real data. This can be done by inspecting the latent space of the model. We compare the positions of the real Rubin \gls{DP1} data in the latent space with the positions of the simulated training data. Any \gls{DP1} targets that are located in low-density regions of the latent space were not well-represented in the training set, and thus their final classifications cannot be trusted. In this work, we use the mean distance to the 10 nearest neighbors in the latent space ($\bar{d}_{10}$) as a density metric. In a similar vein, we also compare the reconstruction loss ($\mathcal{L}_{\rm rec}$) of the real Rubin \gls{DP1} data with that of the simulated training data. If $\mathcal{L}_{\rm rec}$ is worse for any target in the \gls{DP1} data compared to the simulated training set, then the model was not able to reconstruct that target as well as expected, and their final classification is also not reliable. This process is shown in Figure \ref{fig:latent_outliers}, where we plot $\log{(\mathcal{L}_{\rm rec})}$ against $\log{(\bar{d}_{10})}$ of each target in the simulated training set and the real Rubin \gls{DP1} set. We find that the majority of the Rubin \gls{DP1} targets have similar values of $\log{(\mathcal{L}_{\rm rec})}$  and $\log{(\bar{d}_{10})}$ to the simulated training set. This suggests that the simulated templates are representative of the real data and that we can safely use their classifications. 

However, a subset of the \gls{DP1} data have higher values of both $\log{(\mathcal{L}_{\rm rec})}$ and $\log{(\bar{d}_{10})}$, compared to the simulated training data. These are targets whose specific combination of input features were not present in the simulated training set and are effectively out-of-distribution. These can be interpreted as interlopers (e.g., field stars that were not correctly filtered out) or they can also be genuine cluster members that are not \gls{MS} stars or \gls{MS}+\gls{MS} binaries. Alternatively, they could be genuinely new and anomalous targets that are interesting in their own right.\footnote{This is an additional use case of \texttt{flexAE}: finding interesting sources in a population that are \emph{not} well-described by a specific training set. However, this is outside the scope of this work.} Either way, their classifications as single \gls{MS} or \gls{MS}+\gls{MS} binaries with the model trained in Section \ref{sec:training} are not reliable, so for the purposes of this work (i.e., reliably finding binaries in 47 Tucanae), they are discarded from further analysis. In this work, we discard \gls{DP1} targets if their $\log{(\mathcal{L}_{\rm rec})}$ or $\log{(\bar{d}_{10})}$ is higher than the 97$^{\rm th}$ percentile of the simulated training set, which affected \todo{$\sim5.0$\%} of the Rubin \gls{DP1} set. This reduced the sample size to 1,424 targets. These thresholds are shown as dashed lines in Figure \ref{fig:latent_outliers}.

\subsection{Binaries in 47 Tucanae in Rubin DP1}
\label{sec:dp1_binary_47tuc}

\begin{figure}
	\includegraphics[width=\columnwidth]{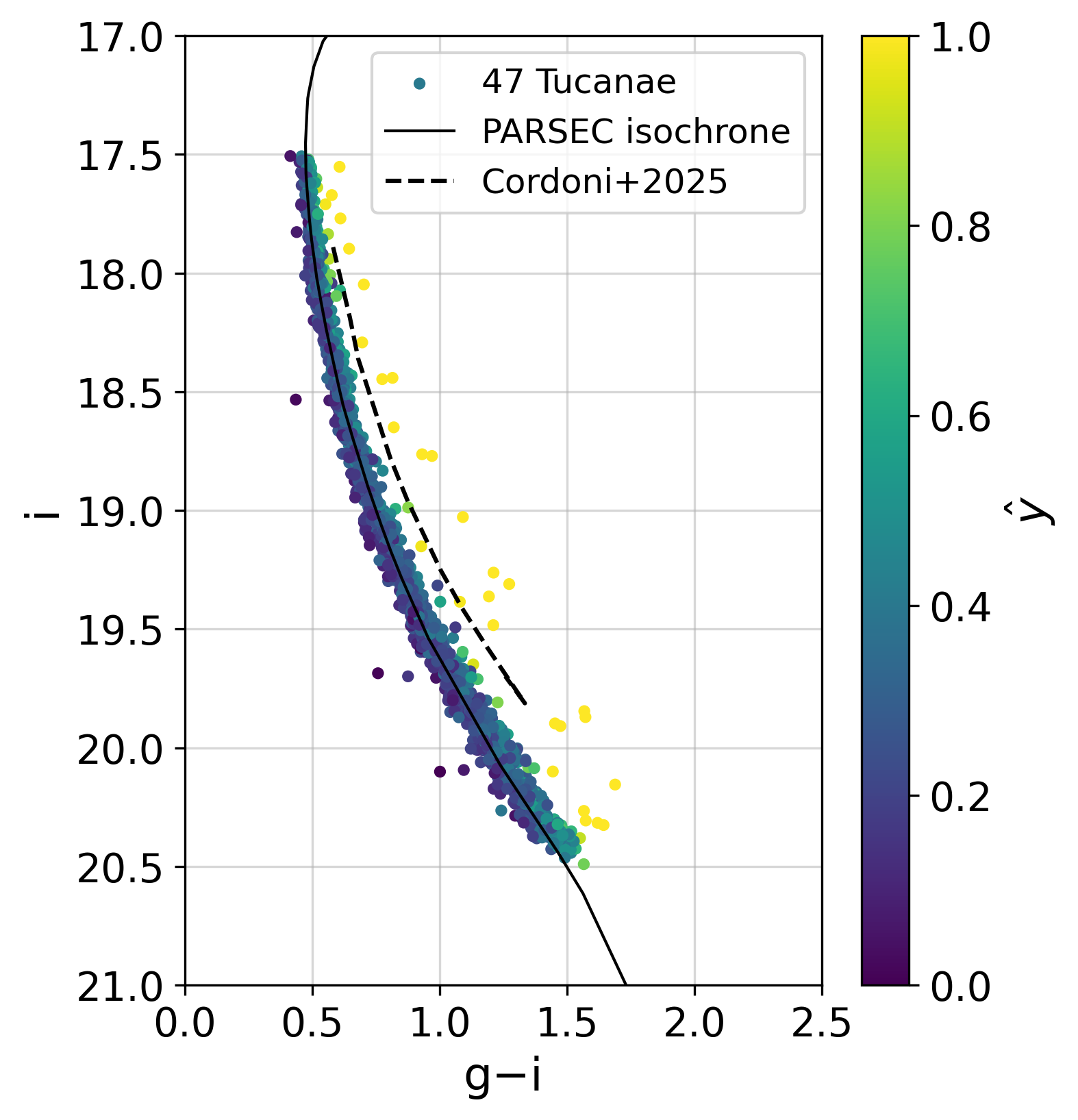}
    \caption{The $i$ vs. $g-i$ \gls{CMD} for the 47 Tuc sample, colored by the probability of the target being a \gls{MS}+\gls{MS} binary system ($\hat{y}$), as predicted by the \gls{AE}. We also show the PARSEC isochrone for 47 Tuc in black and the $q = 0.7$ threshold used by \citet{cordoni_2025} to find \gls{MS}+\gls{MS} binary systems in the $i$ vs. $g-i$ \gls{CMD} as a dashed line. We note that targets closer to the isochrone are more likely to be single star systems (i.e., low $\hat{y}$), while targets with redder colors have a higher probability of being a \gls{MS}+\gls{MS} binary system (i.e., high $\hat{y}$), as expected.}
    \label{fig:cmd_preds}
\end{figure}

We now use the trained \gls{AE} on targets in 47 Tucanae using data from Rubin \gls{DP1} (see Section \ref{sec:sample_selection}) to find candidate \gls{MS}+\gls{MS} binary systems. The class predictions $\hat{y}$ from the model for the 47 Tuc sample are shown on a \gls{CMD} in Figure \ref{fig:cmd_preds}. We also show the PARSEC isochrone for 47 Tuc in black. We note that targets that deviate more towards redder colors from the PARSEC isochrone have higher values for $\hat{y}$, i.e., are considered more likely to be binary systems according to the \gls{AE} model. Keep in mind that this was not explicitly specified, but rather the \gls{AE} inferred this behavior naturally from the training data. This is expected, and is in fact how other \gls{CMD}-based methods are used to find \gls{MS}+\gls{MS} binary stars. We have also plotted the threshold that \citet{cordoni_2025} use to find \gls{MS}+\gls{MS} binary stars in this \gls{CMD}, which corresponds to \gls{MS}+\gls{MS} binary systems with $q = 0.7$. We note that all targets on the red side of the \citet{cordoni_2025} line are confidently classified as binaries by the \gls{AE}. In fact, the mean value of $\hat{y}$ for the \gls{MS}+\gls{MS} binary candidates identified by \citet{cordoni_2025} is equal to \todo{0.998}, while the lowest value of $\hat{y}$ is \todo{0.983}. This suggests a good agreement between both methods. However, we note that we do not explicitly limit ourselves to binaries with $q > 0.7$, which is one of the key advantages of this method, even though low mass ratio \gls{MS}+\gls{MS} binary systems are harder to confidently identify due to the larger difference in flux between the two stars. We also note that, while we are showing the $i$ vs. $g-i$ \gls{CMD} here, the \gls{AE} was trained with information from all available bands (i.e., $gri$).

\begin{figure}
	\includegraphics[width=\columnwidth]{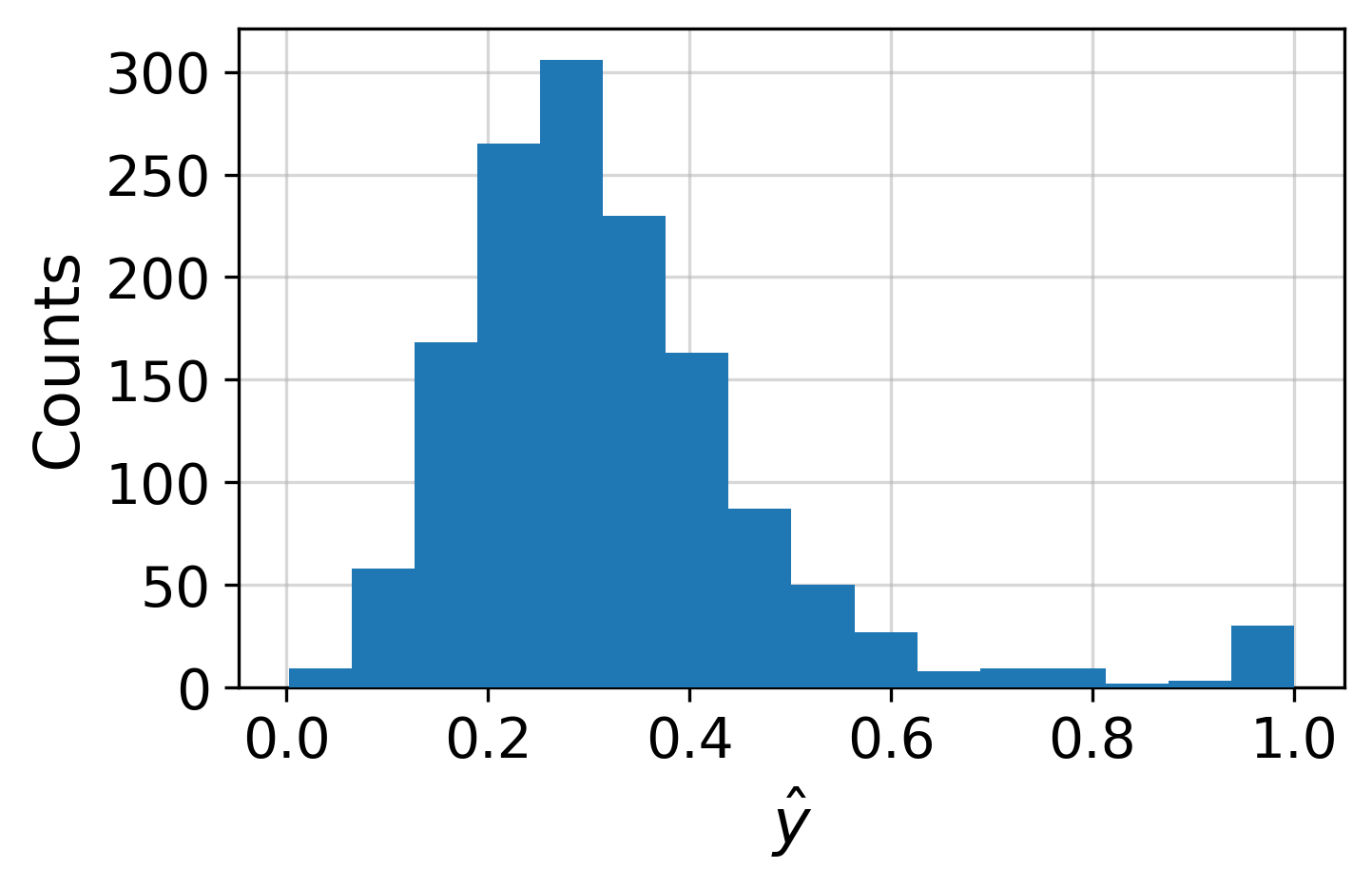}
    \caption{The model class predictions ($\hat{y}$) for the 47 Tucanae targets in Rubin \gls{DP1}. The values of $\hat{y}$ are bimodal: there is a first peak at $\hat{y} \sim 0.3$ and a second, much smaller peak at $\hat{y} \sim 0.95$.}
    \label{fig:yhat_dp1}
\end{figure}

The actual class predictions, $\hat{y}$, are shown in Figure \ref{fig:yhat_dp1}. We find that $\hat{y}$ is bimodal; there is a first broad peak centered at $\hat{y} \sim 0.3$ and a second, albeit much smaller and narrower, peak at $\hat{y} \sim 0.95$. Interestingly, the overall distribution is different from the distribution of $\hat{y}$ for the simulated set (see Figure \ref{fig:classification}). This is not due to out-of-distribution targets that appear in the \gls{DP1} set but not in the simulated training set, as these are filtered out in Section \ref{sec:latent_space_outliers}. Instead, we suspect that this is caused by two different reasons. Firstly, the peak of targets at $\hat{y} \sim 0.95$ (i.e., very confidently classified as binary) is lower in the distribution for the real \gls{DP1} data in Figure \ref{fig:yhat_dp1} compared to the corresponding peak in the distribution for the simulated targets shown in Figure \ref{fig:classification}. This is likely caused by a mismatch in the relative proportions of single stars and binary systems between the simulated training set and the real data. The simulated set contained 50\% single stars and 50\% binary systems, while we expect fewer binary systems in the real data. Secondly, the \gls{DP1} distribution lacks a strong and narrow peak of objects at $\hat{y} \sim 0.0$ that was seen in the training data. As noted in Section \ref{sec:creating_training_set}, we varied the luminosity by $\pm0.2$ dex to increase the diversity in the training set. However, this caused a lot of the simulated single stars in the training set to appear below the isochrone (i.e., on the blue side) in the $i$ vs. $g-i$ \gls{CMD}. These simulated targets are confidently classified as single stars and have low values of $\hat{y}$. However, analogous systems are almost entirely absent from the real \gls{DP1} data for 47 Tucanae. Indeed, when excluding simulated targets that appear below the isochrone in the $i$ vs. $g-i$ \gls{CMD} from the analysis, this peak disappears almost entirely (see Appendix \ref{app:yhat_distribution} for more details).


\begin{figure}
	\includegraphics[width=\columnwidth]{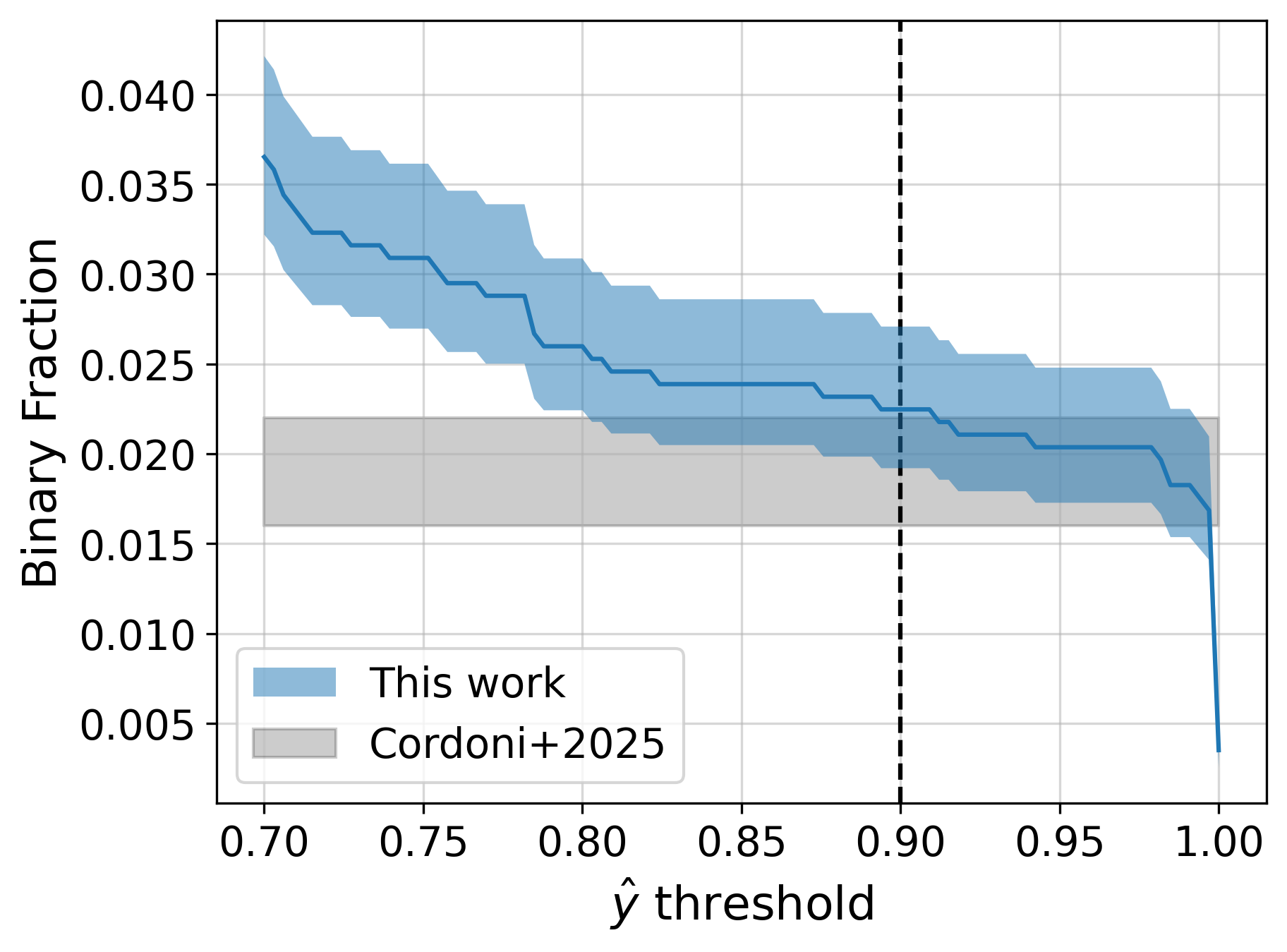}
    \caption{The \gls{MS}+\gls{MS} binary fraction plotted against the $\hat{y}$ threshold used to classify systems as binary. The blue shaded region represents statistical uncertainties on the binary fraction, estimated using the beta distribution quantile technique \citep{cameron_2011}. The black shaded region represents the different \gls{MS}+\gls{MS} binary fractions found by \citet{cordoni_2025}, which range between $1.6 - 2.2$\% depending on which \gls{CMD} was used. The \gls{MS}+\gls{MS} binary fraction decreases as stricter (i.e., higher $\hat{y}$) thresholds are used. The binary fraction is \todo{$2.2^{+0.5}_{-0.3}\%$} when using \todo{$\hat{y} > 0.9$}.}
    \label{fig:binary_frac}
\end{figure}

However, this means that we cannot simply use the standard choice of $\hat{y} = 0.5$ as a threshold to distinguish between single stars and binary systems, as that would classify the tail end of the first peak as binary systems. Instead, we will use a higher and more conservative threshold. The effect of the $\hat{y}$ threshold on the \gls{MS}+\gls{MS} binary fraction is shown in Figure \ref{fig:binary_frac}. We find that the binary fraction decreases as stricter (i.e., higher $\hat{y}$) thresholds are used. The \gls{MS}+\gls{MS} binary fraction equals \todo{$3.6\%$} when using $\hat{y} > 0.7$ as a threshold to select all binaries. However, it seems likely that this threshold still includes the tail end of the first peak and thus may be contaminated with a number of single stars. The \gls{MS}+\gls{MS} binary fraction seems to flatten out around $\hat{y} \sim 0.85 - 0.95$ with values between \todo{$2.0 - 2.4 \%$}. Based on Figures \ref{fig:yhat_dp1} and \ref{fig:binary_frac}, we decide to classify all targets with \todo{$\hat{y} > 0.9$} as \gls{MS}+\gls{MS} binary systems. This results in an observed \gls{MS}+\gls{MS} binary fraction of \todo{$2.2^{+0.5}_{-0.3}\%$} in 47 Tucanae. The uncertainties represent statistical uncertainties calculated with the beta distribution quantile technique \citep{cameron_2011}. We note that \todo{$\hat{y} > 0.9$} is a conservative threshold that removes all targets that the \gls{AE} is uncertain about and removes false positives, at the cost of completeness (especially at lower mass ratios), and thus effectively acts as a lower limit to the binary fraction. The real \gls{MS}+\gls{MS} binary fraction (which includes lower mass ratios) is likely higher, which is discussed in more detail in Section \ref{sec:correction_binary_fraction}.

A \gls{MS}+\gls{MS} binary fraction of \todo{$2.2^{+0.5}_{-0.3}\%$} in the outskirts of 47 Tuc is in good agreement with the results of \citet{cordoni_2025}, who use the same Rubin \gls{DP1} data to look for binary systems in 47 Tuc. They find \gls{MS}+\gls{MS} binary fractions of $1.6\pm0.5$\%, $1.8\pm0.6$\%, and $2.2\pm0.7$\% for systems with $q>0.7$ when using an $i$ vs. $g-i$ \gls{CMD}, $i$ vs. $r-i$ \gls{CMD} and $r$ vs. $g-r$ \gls{CMD}, respectively.

We want to highlight that we do not explicitly limit our sample to any mass ratios. However, given our agreement with \citet{cordoni_2025}, who did limit themselves to binary systems with $q>0.7$, we suspect that the trained model used in this work is most sensitive to $q>0.7$ as well. This is consistent with the results from Figure \ref{fig:completeness_2d}, where we found that the completeness in the simulated test set decreases for binary systems with large differences between the temperatures of the primary and secondary stars. It seems that the three $gri$ bands alone are not enough to find binaries with lower mass-ratios. This will be verified in a companion paper (G\'eron et al. in preparation), where we will characterize the targets found in this work in more detail and measure their temperatures and mass ratios through broadband \gls{SED} fitting. We expect to be more sensitive to low mass-ratio binaries as more bands are included in the analysis as they become available with future data releases of Rubin \gls{LSST}. The redder $z$ and $y$ bands will be especially helpful to identify low mass-ratio binaries, as the cooler secondary star will emit more flux at longer wavelengths.

\begin{figure}
    \centering
	\includegraphics[width=\columnwidth]{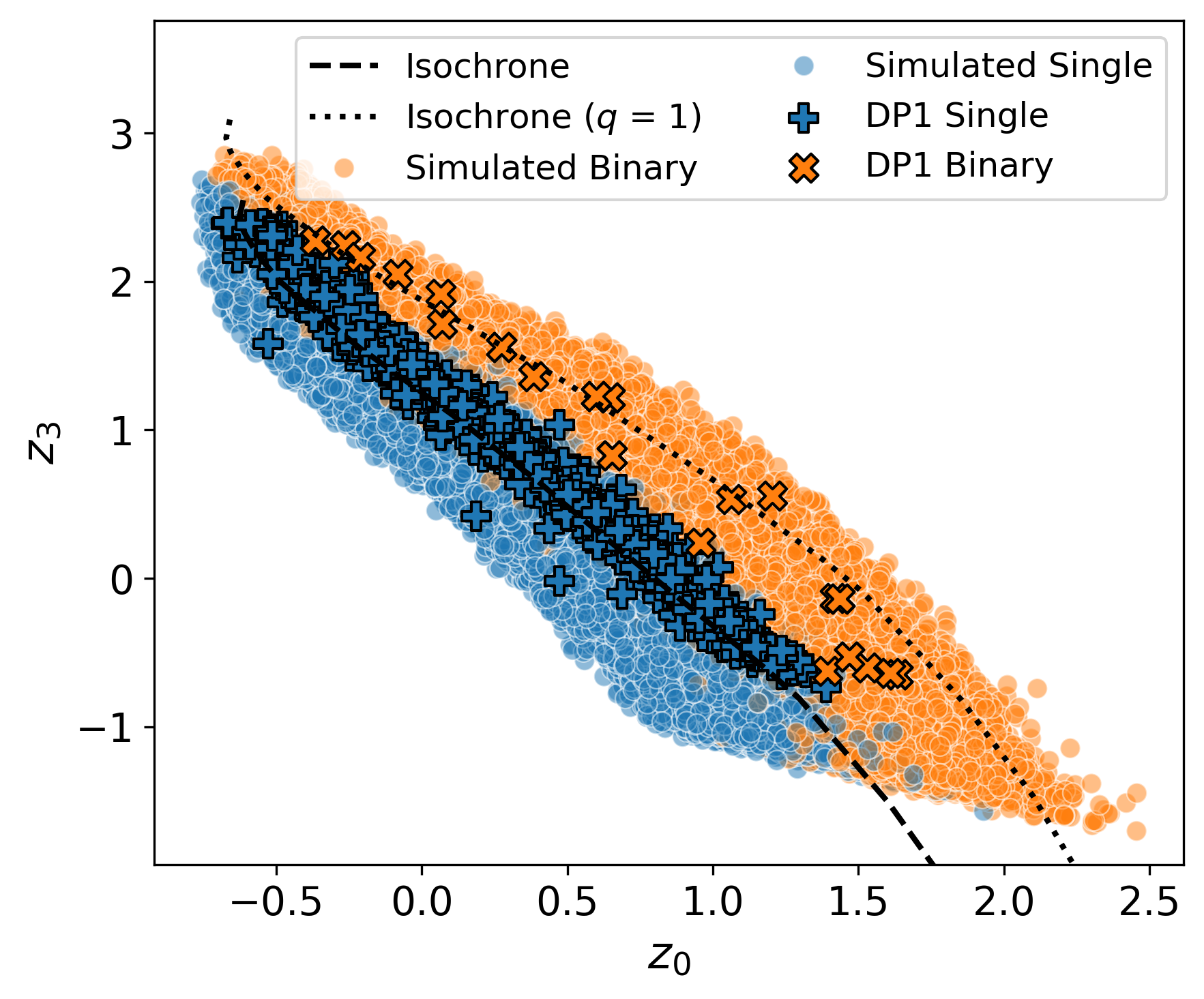}
    \caption{The positions of the single stars and \gls{MS}+\gls{MS} binary systems found in 47 Tucanae with data from Rubin DP1 (the blue plus signs and orange crosses, respectively) in two dimensions of the latent space. We also plot the single stars and \gls{MS}+\gls{MS} binary systems in the simulated training set (blue and orange circles, respectively), as well as the main sequence PARSEC isochrone (dashed black line) and the $q=1$ isochrone (dotted line).}
    \label{fig:latent_space_dp1}
\end{figure}

Finally, the positions of the identified single stars and \gls{MS}+\gls{MS} binary systems in Rubin \gls{DP1} in two of the latent space dimensions are shown in Figure \ref{fig:latent_space_dp1}. As expected, the targets classified as single stars are located in the same area of the latent space as the single stars in the simulated training set. They also cluster around the main sequence isochrone. In contrast, the targets identified as \gls{MS}+\gls{MS} binary systems in Rubin \gls{DP1} are confidently located in the region of the latent space occupied by the simulated \gls{MS}+\gls{MS} binary systems and are closer to the $q=1$ isochrone.

\subsection{Correcting the Observed Binary Fraction}
\label{sec:correction_binary_fraction}

We can make an attempt at correcting the observed \gls{MS}+\gls{MS} binary fraction and estimating the true \gls{MS}+\gls{MS} binary fraction using the precision and completeness values estimated from the simulated test set (see Figure \ref{fig:precision_completeness}). With a threshold of $\hat{y} > 0.9$, the precision and completeness equal \todo{0.98} and \todo{0.60}, respectively. Using these values to correct the observed \gls{MS}+\gls{MS} binary fraction implies a true \gls{MS}+\gls{MS} binary fraction of \todo{$3.7^{+0.8}_{-0.5}\%$}. However, we note that it is not obvious that we can simply use the total completeness and precision values from the simulated test set. This is because, first of all, the true correction is also temperature-dependent (see Figure \ref{fig:completeness_2d}). Secondly, the simulated templates will inevitably differ from the real data in subtle ways. Finally, the simulated set was constructed with a broader range of metallicities, surface gravities, and $A_V$ values than typically found in 47 Tucanae (see Section \ref{sec:sample_selection}). 

Alternatively, if we are currently indeed only sensitive to binaries with $q>0.7$ (see Section \ref{sec:dp1_binary_47tuc}) and assume a flat mass ratio distribution, we can extrapolate the observed \gls{MS}+\gls{MS} binary fraction of \todo{$2.2^{+0.5}_{-0.3}\%$} to a total \gls{MS}+\gls{MS} binary fraction of $\todo{7.5^{+1.5}_{-1.1}}$\%. 

Both of these approaches are simple extrapolations. However, they result in useful approximations of the total \gls{MS}+\gls{MS} binary fraction, which probably lies somewhere between \todo{$3.7^{+0.8}_{-0.5}\%$} and $\todo{7.5^{+1.5}_{-1.1}}$\%. We will better quantify the completeness correction in a companion paper using \texttt{MultiStarFitting} to allow for a temperature-dependent correction.

\subsection{Radial Dependence of the Binary Fraction in 47 Tucanae}



\begin{figure}
	\includegraphics[width=\columnwidth]{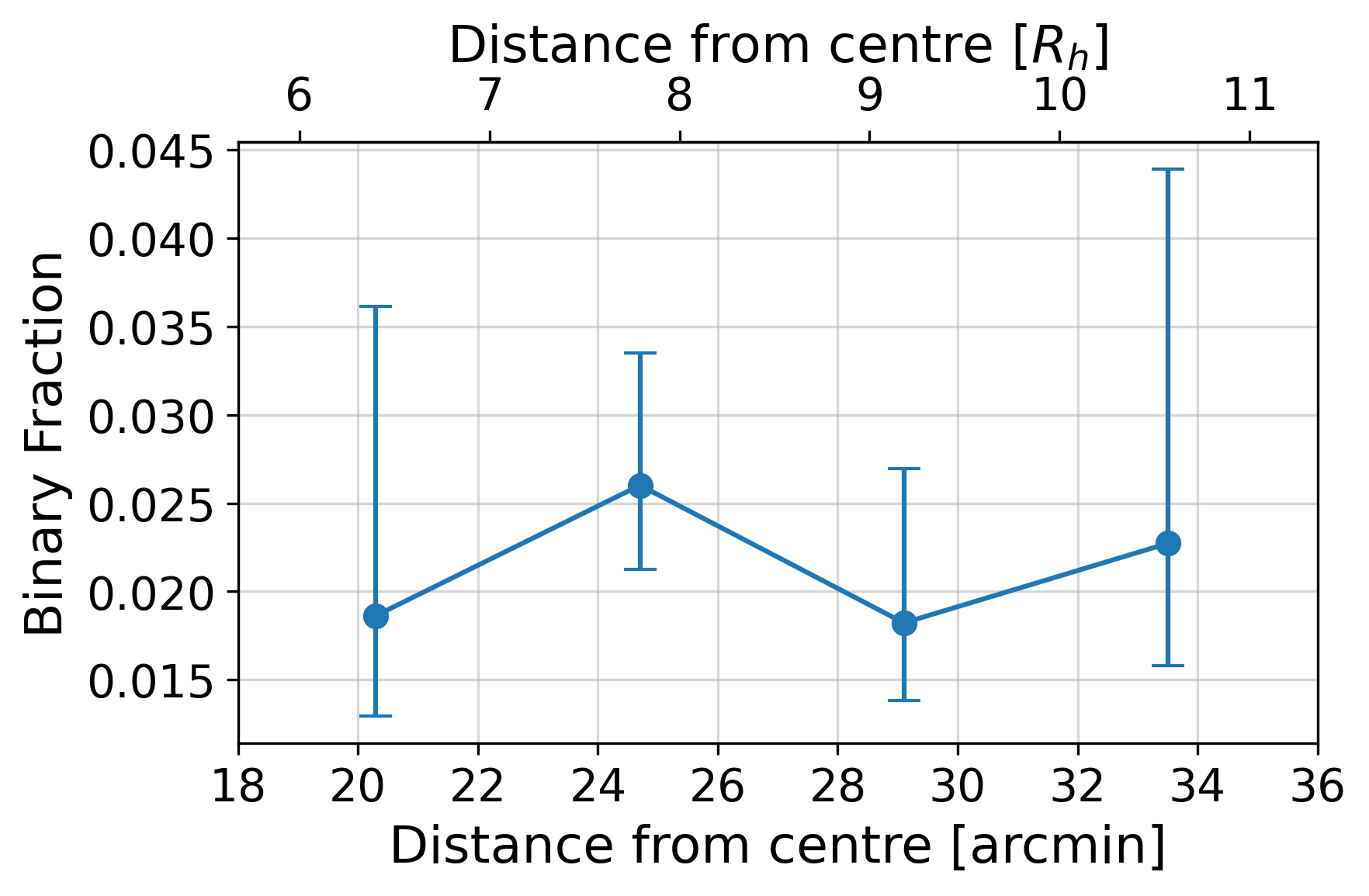}
    \caption{The \gls{MS}+\gls{MS} binary fraction at different distances from the cluster center in terms of arcmin (bottom x-axis) and cluster half-light radii (top x-axis), assuming a half-light radius of 3.17 arcmin \citep{harris_1996}. We observe no statistically significant trend. The error bars represent statistical uncertainties calculated with the beta distribution quantile technique \citep{cameron_2011}.}
    \label{fig:radial_binary_frac}
\end{figure}

The Rubin \gls{DP1} sample contains targets in the outskirts of 47 Tucanae, at distances of $\sim18-36$ arcmin (or $\sim5.7 -11.4$ $R_h$) from the cluster center. The radial dependence of the \gls{MS}+\gls{MS} binary fraction in that range is shown in Figure \ref{fig:radial_binary_frac}. The \gls{MS}+\gls{MS} binary fraction is constant within that distance range, as no statistically significant trend is found.

The binary fraction in globular clusters is known to decrease with radius from the cluster center \citep{sollima_2007, milone_2012}. The overall binary fraction in globular clusters and its radial trend is determined by two main factors. First of all, the dense cluster environment disrupts binary systems, which results in significantly lower binary fractions in globular clusters compared to field stars \citep{cool_2002, ivanova_2005, sollima_2007}. Secondly, mass segregation causes binary systems to pile up in the center of the cluster \citep{goodman_1989,heggie_2006}. For 47 Tuc specifically, \citet{ji_2015} find a binary fraction of $4.14\pm0.29$\% for systems with $q>0.5$ in the core of the cluster (0.00 - 0.29 $R_h$). This decreases to $2.72\pm0.22$\% and $2.11\pm0.18$\% as the radius increases to 0.29-0.42 $R_h$ and 0.42-0.56 $R_h$, respectively. \citet{milone_2012} find a \gls{MS}+\gls{MS} binary fraction for 47 Tuc of $0.5\pm0.3$\% for systems with $q>0.7$ and $0.9\pm0.3\%$ for systems with $q>0.5$ between the core and half-mass radius. \citet{muller_horn_2025} also find a radial dependence on the binary fraction for 47 Tuc. The binary fraction is $\sim5.5$\% in the core of the cluster, which then decreases to $\sim2-4$\% between the core and half-light radius. Note that their analysis includes binary systems with all mass ratios and all types of binaries, not just \gls{MS}+\gls{MS} binaries.

As noted above, our observed \gls{MS}+\gls{MS} binary fraction is \todo{$2.2^{+0.5}_{-0.3}\%$} between $\sim5.7-11.4$ $R_h$, while the true total \gls{MS}+\gls{MS} binary fraction is likely between \todo{$3.7^{+0.8}_{-0.5}\%$} and $\todo{7.5^{+1.5}_{-1.1}}$\%. This is higher than the aforementioned binary fractions between the core and half-mass radius measured by \citet{milone_2012, ji_2015, muller_horn_2025}, suggesting that the binary fraction rises again in the outskirts of 47 Tuc. Interestingly, \citet{muller_horn_2025} also find that the binary fraction for 47 Tuc increases again to $\sim8$\% at the half-light radius, though they argue that this is possibly a spurious measurement due to the low number of pointings in their sample. However, based on the findings presented in this work, we suggest that their observation was not spurious. \citet{cordoni_2025} also concluded that the binary fraction in the outskirts of 47 Tuc is higher than the fractions reported for the inner region of the cluster. We agree with their interpretation that the lower stellar density in the cluster outskirts results in fewer encounters. Thus, binary systems are less likely to be disrupted and the binary fraction remains higher. \citet{cadelano_2026} also find this bimodal behavior for the binary fraction using a sample of six different globular clusters, which is explained by \citet{bruce_2026} as a combination of mass segregation and binary disruption. We will investigate the binary fraction in the outskirts of more globular clusters with Rubin \gls{LSST} in future work, and determine whether this behavior is observed in additional globular clusters as well.

\section{Conclusion}
\label{sec:conclusion}

In this work, we introduced \texttt{flexAE}, a novel and adaptable \gls{AE} architecture designed to identify unresolved binary systems using multi-band photometry. This framework is, in simple terms, essentially a multi-dimensional \gls{CMD}. However, we are able to use information from all available bands simultaneously and explicitly account for observational uncertainties. Furthermore, the \gls{AE} finds the most efficient way to classify single stars and binary systems in this high-dimensional parameter space, as opposed to following an isochrone. It also organizes its latent space in such a way that is optimal for classification by training a classifier component at the same time as the encoder and decoder components. Another main advantage of this approach is that, in principle, we could be sensitive to binary systems with lower mass-ratios (and thus lower flux ratios), though it depends on the quality of the photometry and wavelength coverage. Finally, it is trivial to add additional bands to the analysis as they become available.

We use \texttt{flexAE} to find \gls{MS}+\gls{MS} binaries in the outskirts of 47 Tucanae. We trained and validated the framework with a simulated dataset of single star and binary systems created with the stellar atmosphere templates from \citet{castelli_2003} convolved with the throughput curves of Rubin \gls{DP1}. The model was able to robustly reconstruct the input features, achieved a classification accuracy of \todo{0.85} on the simulated test set, and demonstrated a strongly bimodal prediction probability ($\hat{y}$). We also showed that we can use the latent space of the model to identify anomalies and out-of-distribution targets by using the mean distance to the 10 nearest neighbors in the latent space ($\bar{d}_{10}$) alongside the reconstruction loss ($\mathcal{L}_{\rm rec}$).

We then applied the model to 47 Tucanae cluster members found in Rubin \gls{DP1}. We adopt a conservative classification threshold of \todo{$\hat{y} > 0.9$} to minimize single star contamination and find an observed \gls{MS}+\gls{MS} binary fraction of \todo{$2.2^{+0.5}_{-0.3}\%$} in the outskirts of 47 Tucanae, which is in general agreement with the literature (e.g., \citealt{cordoni_2025}). We emphasize that this value is a lower limit, as it is likely that lower mass ratio binary systems were missed. Using simplified extrapolations, we conclude that the intrinsic total binary fraction lies between \todo{$3.7^{+0.8}_{-0.5}\%$} and $\todo{7.5^{+1.5}_{-1.1}}$\%. 

We also note that the observed binary fraction stays constant between 18 - 36 arcmin (or \todo{$5.7-11.4$} half-light radii) from the cluster center. Comparison to the literature suggests a non-monotonic binary fraction in 47 Tucanae as a function of radius; it is high in the core of the cluster and decreases with radius until approximately the half-light radius. Our work shows that the binary fraction increases again in the outskirts of 47 Tucanae, well beyond the half-light radius, forming an overall U-shaped trend.

We will analyze the binary systems found in this paper in greater detail in a companion paper, where we will use \texttt{MultiStarFitting} to characterize the systems, determine their mass ratios, and derive a more robust correction to obtain the intrinsic binary fraction.

The flexibility of the \texttt{flexAE} architecture makes it an ideal tool for large-scale, automated binary star classification in the era of wide-field photometric surveys. This work represents a first step towards preparing for upcoming data releases from the Rubin Observatory, such as \gls{DP2} and \gls{DR1}, which will enable us to find more binary systems in more clusters at a much larger scale. As more data from Rubin \gls{LSST} becomes available in \gls{DR1} and 47 Tucanae is imaged with more bands, we will add them to the analysis. We expect that using colors from all six $ugrizy$ bands will enable us to find \gls{MS}+\gls{MS} binary systems with even lower mass ratios. Finally, we want to emphasize that \texttt{flexAE} is adaptable and independent of any specific set of filters or choice of stellar library. It can easily be modified to find other types of binary systems with other surveys (e.g., see Appendix \ref{app:YSG} for an example of how to use \texttt{flexAE} to find \gls{YSG}+\gls{MS} binaries).

The \texttt{flexAE} code is publicly available on GitHub\footnote{\url{https://github.com/tobiasgeron/flexAE}}, together with the trained model and the full table of the \gls{DP1} binaries found in this work.

\begin{acknowledgments}

T.G. is a Canadian Rubin Fellow at the Dunlap Institute. The Dunlap Institute is funded through an endowment established by the David Dunlap family and the University of Toronto. T.G. also is supported through the LSST-DA Catalyst Fellowship; this publication was thus made possible through the support of grant 62192 from the John Templeton Foundation to LSST-DA.

A.L. acknowledges support from the NSERC and is funded through a NSERC Canada Graduate Scholarship—Doctoral. A.L. is also supported by the Data Sciences Institute at the University of Toronto through grant number DSIDSFY3R1P02. 

J.S.S. was supported by NSERC Discovery Grant RGPIN-2023-04849.

M.R.D. acknowledges support from NSERC through grant RGPIN-2025-06224, the Canada Research Chairs Program (CRC-2023-00127), the Ontario ERA program (ER22-17-164) and the Dunlap Institute at the University of Toronto.

This material is based upon work supported in part by the National Science Foundation through Cooperative Agreements AST-1258333 and AST-2241526 and Cooperative Support Agreements AST-1202910 and 2211468 managed by the Association of Universities for Research in Astronomy (AURA), and the Department of Energy under Contract No. DE-AC02-76SF00515 with the SLAC National Accelerator Laboratory managed by Stanford University. Additional Rubin Observatory funding comes from private donations, grants to universities, and in-kind support from LSST-DA Institutional Members.

This research uses services or data provided by the Rubin Science Platform at NSF-DOE Vera C. Rubin Observatory, which is jointly funded by the U.S. National Science Foundation and the U.S. Department of Energy, Office of Science.

This work has made use of data from the European Space Agency (ESA) mission
{\it Gaia} (\url{https://www.cosmos.esa.int/gaia}), processed by the {\it Gaia}
Data Processing and Analysis Consortium (DPAC,
\url{https://www.cosmos.esa.int/web/gaia/dpac/consortium}). Funding for the DPAC
has been provided by national institutions, in particular the institutions
participating in the {\it Gaia} Multilateral Agreement.

\end{acknowledgments}

\software{\texttt{Astropy} \citep{astropy_2013, astropy_2018, astropy_2022}, \texttt{Matplotlib} \citep{matplotlib_2007}, \texttt{NumPy} \citep{numpy_2020}, \texttt{SciPy} \citep{scipy_2020}, \texttt{PyTorch} \citep{pytorch_2019}, \texttt{coordinate\_matching}\footnote{\url{https://github.com/tobiasgeron/coordinate_matching}}, \texttt{flexAE}\footnote{\url{https://github.com/tobiasgeron/flexAE}}, \texttt{MultiStarFitting}}



\appendix

\section{Training \texttt{flexAE} to find Yellow Supergiant Binaries}
\label{app:YSG}

The focus of this paper was to use \texttt{flexAE} to identify \gls{MS}+\gls{MS} binaries in 47 Tucanae with photometry from Rubin \gls{DP1}. However, \texttt{flexAE} can easily be adapted to find other types of binary systems (or even do other astronomical classification tasks). To illustrate this point, we will use \texttt{flexAE} to train a model that is able to identify \gls{YSG} binaries. 

As noted by \citet{ogrady_2024}, the binary fraction of \glspl{YSG} in the Hertzsprung gap is not well constrained. They use color-color diagrams to identify binary systems with a \gls{YSG} and O- or B-type \gls{MS} companion. They are able to identify hundreds of candidate \gls{YSG} binary systems in the \gls{LMC} and \gls{SMC}. Here, we will attempt to train a model to find these same types of \gls{YSG} binaries. However, Rubin \gls{LSST} will be saturated at $r \sim 16$, which means that \glspl{YSG} in the Magellanic Clouds will be too bright to study with Rubin. Instead, we create a simulated sample of targets placed at random distances between \todo{500} - \todo{2,000} kpc. We use the stellar atmosphere model templates of \citet{castelli_2003} to create a sample of single \glspl{YSG} and binary systems that have one \gls{YSG} and an O- or B-type \gls{MS} companion. We assume that the magnitudes are already corrected for extinction. We follow the definition of \citet{ogrady_2024} for a \gls{YSG}: $\log(L/L_\odot) \geq 4.0$, effective temperatures between $4,000 \textrm{ K}\leq T_{\rm eff} \leq 9,000 \textrm{ K}$, and $M_* \geq 9 M_\odot$. We also enforce that the minimum temperature of the companion must be greater than $11,000$ K to ensure O- or B-type \gls{MS} companions. Finally, we include templates with all metallicities ($-2.5 < [M/H] < 0.5$) and all surface gravities ($0.0 < \log(g \;[\rm{cm\;s^{-2}}]) < 5.0$) that are available in the \citet{castelli_2003} models to increase the diversity in the training set. 

We assume that we have accurate photometry in all the six Rubin bands ($ugrizy$) and we create the following colors that capture the shape of the broadband \gls{SED}: $u-g$, $g-r$, $r-i$, $i-z$, and $z-y$. We also use the absolute magnitude in the $r$-band ($M_r$) to anchor the vertical scaling. We choose the absolute magnitude, as opposed to the apparent magnitude, as our targets are placed at a range of distances. We also use their associated uncertainties ($\sigma_{M_r}$, $\sigma_{u-g}$, $\sigma_{g-r}$, $\sigma_{r-i}$, $\sigma_{i-z}$, and $\sigma_{z-y}$) as input features for the error head. This results in a total of 12 input features for the \gls{AE}. We use the same model architecture described in Section \ref{sec:ae_architecture}, broadly follow the steps and parameters discussed in Section \ref{sec:creating_training_set} to create the simulated training set, and train the model as described in Section \ref{sec:training}.

\begin{figure}
	\includegraphics[width=0.8\columnwidth]{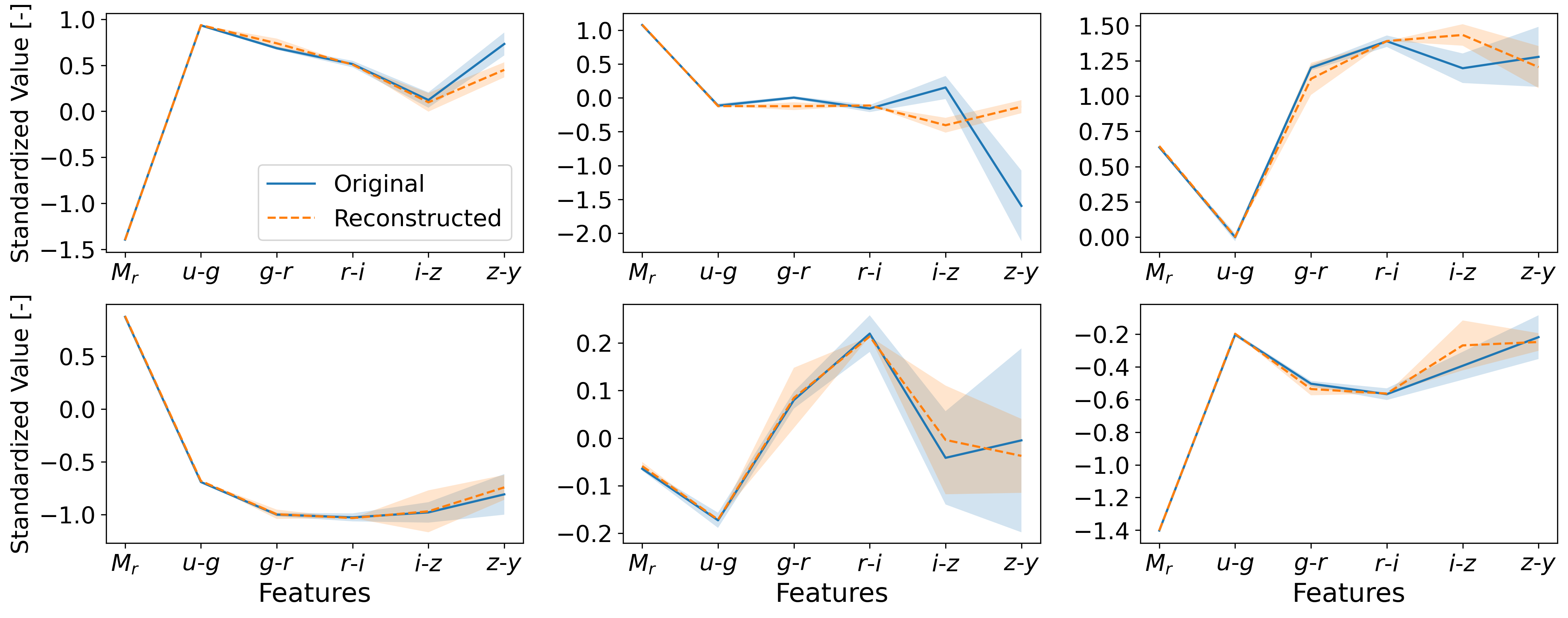}
    \centering
    \caption{A comparison of the standardized input features (blue) as well as the reconstructed \gls{AE} output features (orange) for randomly selected targets in the training set. The shaded blue regions represent the input uncertainties, whereas the shaded orange region represents the scatter for each feature. The \gls{AE} reconstruction is similar to the input.}
    \label{fig:YSG_training_reconstructed_features}
\end{figure}

\begin{figure}
	\includegraphics[width=0.4\columnwidth]{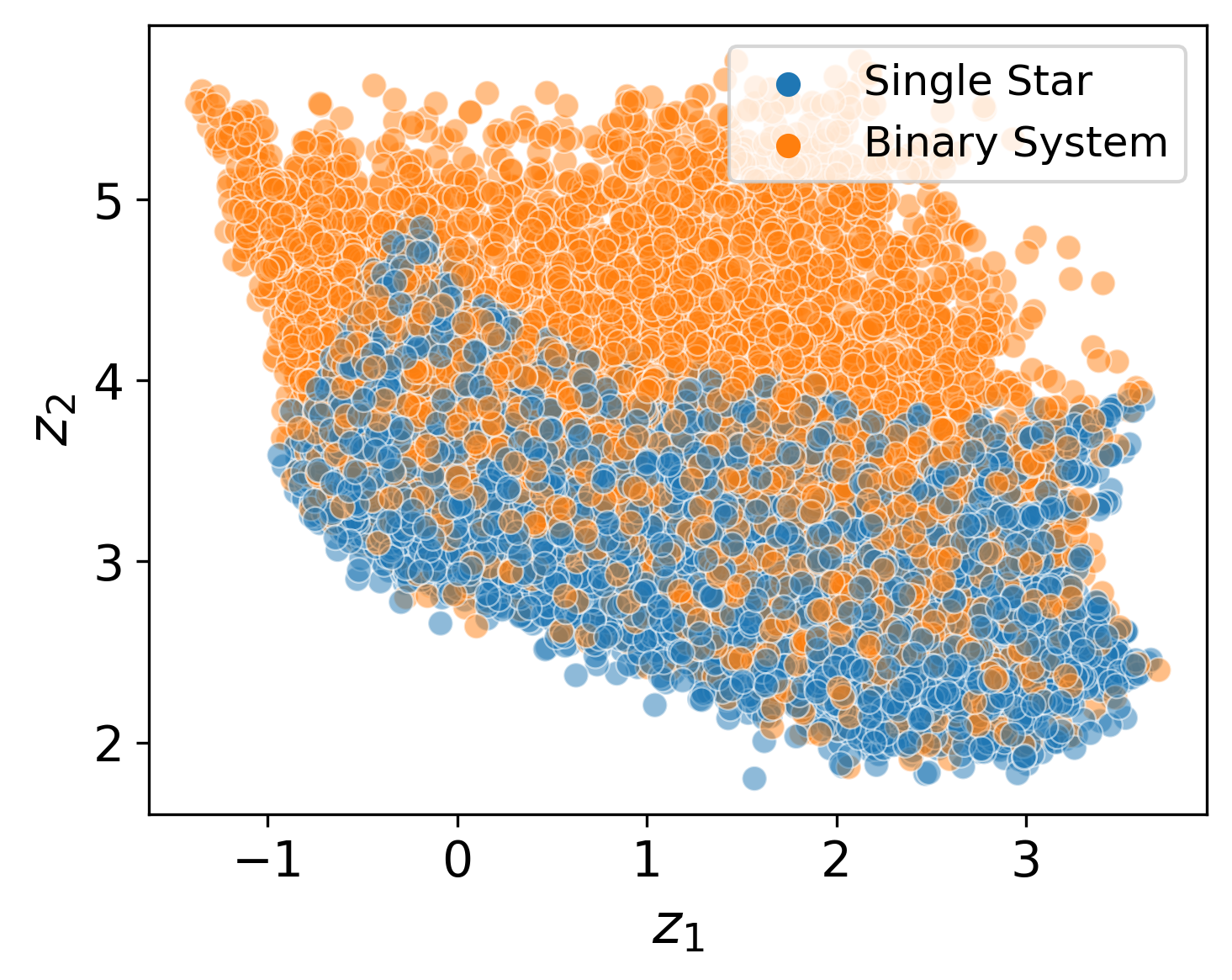}
    \centering
    \caption{The location of the single stars (blue) and binary systems (orange) in the training sample in the \todo{$z_1$} vs. \todo{$z_2$} latent dimensions. There is some overlap in the latent space for both classes, but there is a population of binary systems that are clearly separated in the top half (high \todo{$z_2$}) of this plot.}
    \label{fig:YSG_latent_space}
\end{figure}

The input features and their reconstructions are shown in Figure \ref{fig:YSG_training_reconstructed_features}. The model is able to reconstruct the first four features ($M_r$, $u-g$, $g-r$, $r-i$) well, but struggles more with the latter two features ($i-z$ and $z-y$). However, these are also the features with the largest relative uncertainty, as shown by the blue shaded regions in Figure \ref{fig:YSG_training_reconstructed_features}. Increasing the number of latent space dimensions might improve reconstruction even more. However, we believe that the reconstruction was relatively successful, which implies that the model learned to store the relevant information for reconstruction in the four-dimensional latent space. The locations of the single \gls{YSG} stars and the binary systems with one \gls{YSG} and O- or B-type \gls{MS} companion in the \todo{$z_1$} vs. \todo{$z_2$} latent dimensions are shown in Figure \ref{fig:YSG_latent_space}. There is some overlap between the two classes in the lower half (low \todo{$z_2$}) of the plot. However, some binary systems are clearly separated in the top half (high \todo{$z_2$}) of this plot.

\begin{figure}
	\includegraphics[width=0.7\columnwidth]{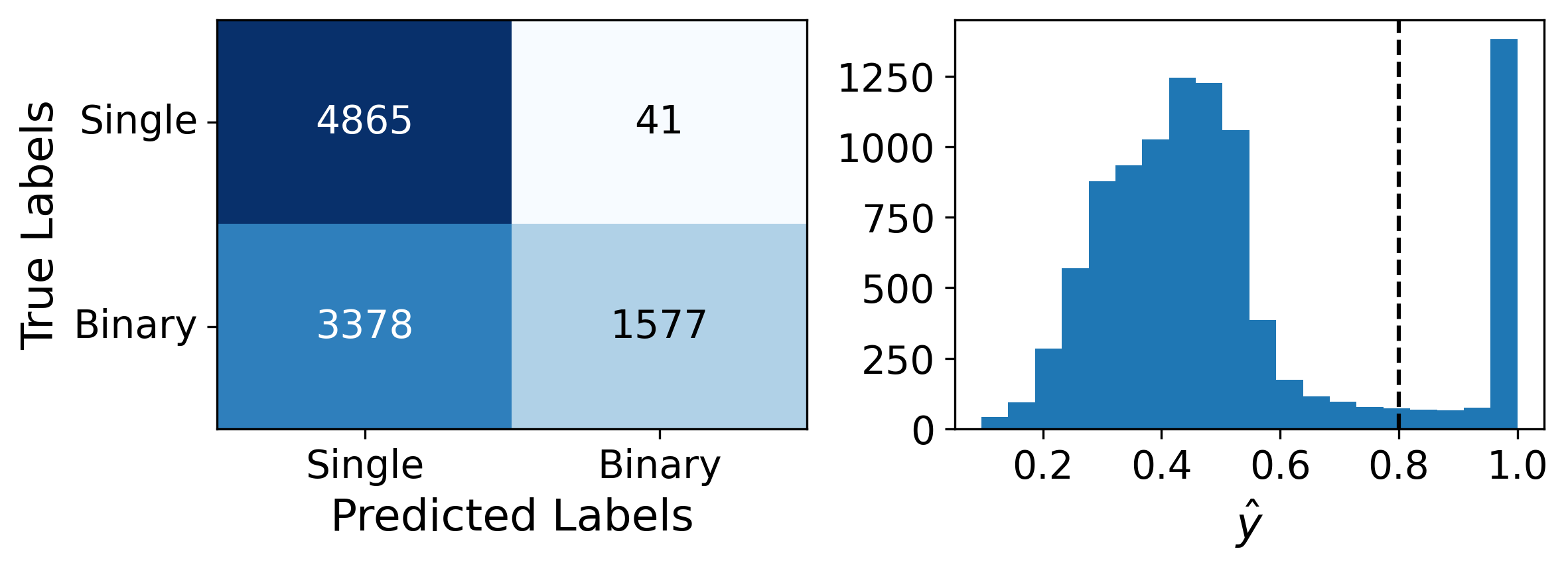}
    \centering
    \caption{The left panel shows the confusion matrix of the real and predicted classes in the test set. A threshold of $\hat{y} = 0.8$ was used to distinguish between single stars and binary systems. The full distribution of $\hat{y}$ for the test set is shown in the right panel.}
    \label{fig:YSG_classification}
\end{figure}

\begin{figure}
	\includegraphics[width=0.5\columnwidth]{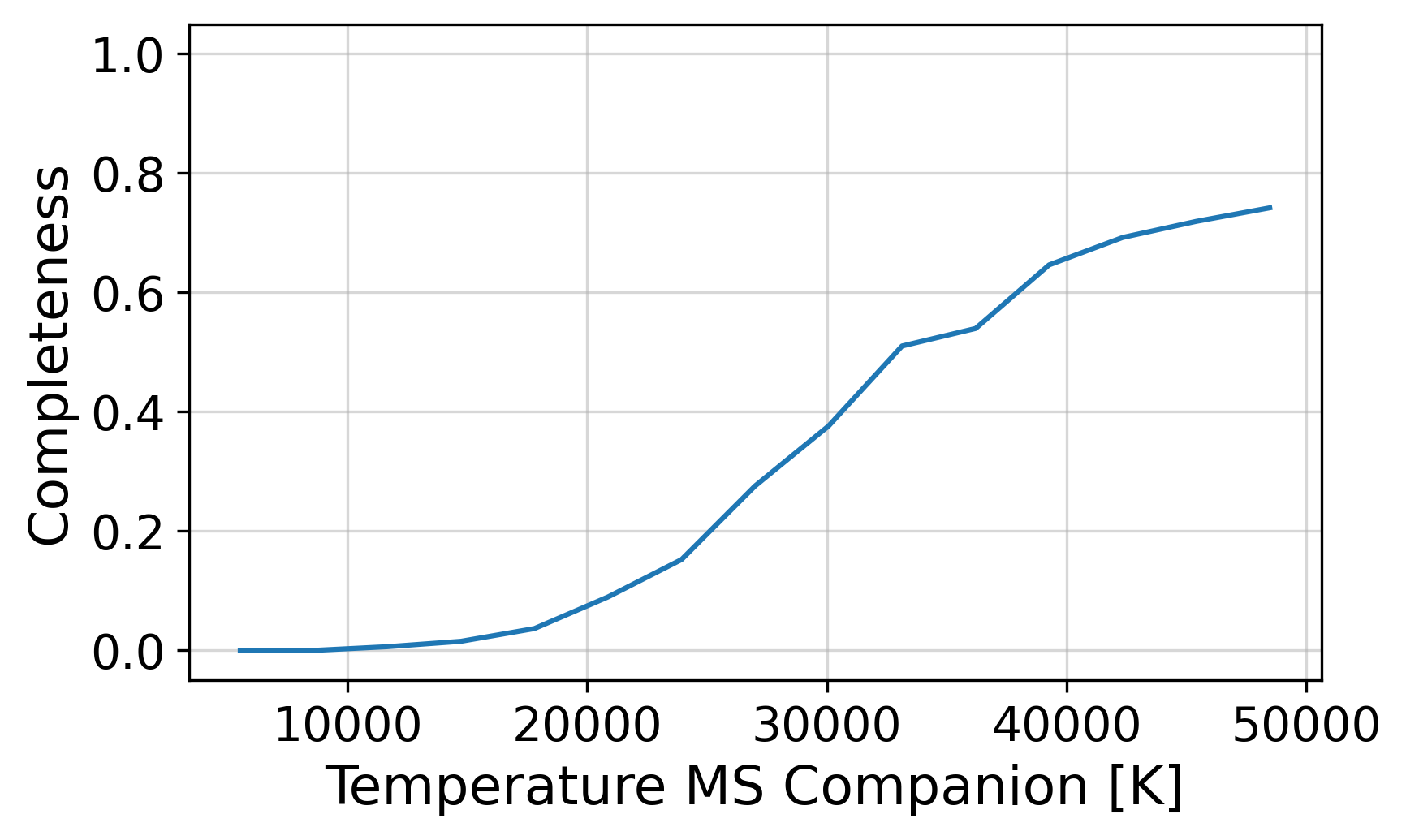}
    \centering
    \caption{The completeness of the binary class plotted against the temperature of the \gls{MS} companion of the binary. The completeness is near 0.0 when the companion temperature is around 10,000 K (i.e., cooler B-type stars), but rises up to $\sim0.7$ as the temperature gets closer to 50,000 K (i.e., hot O-type stars).}
    \label{fig:YSG_completeness_companion_temperature}
\end{figure}

The model performance for classification on the simulated test set is shown in Figure \ref{fig:YSG_classification}. The right panel shows the full distribution of the model class predictions, $\hat{y}$. The distribution has two clearly separate peaks, one around $\hat{y} \sim 0.45$ and a second around $\hat{y} \sim 0.95$. It seems that the model is very confident that a large number of targets are binary systems, which likely corresponds to the binary systems that were clearly separated from the single stars in the latent space in Figure \ref{fig:YSG_latent_space}. We decide to use a threshold of $\hat{y} = 0.8$ to distinguish between single stars and binary systems. A confusion matrix is shown in the left panel of Figure \ref{fig:YSG_classification}. As expected, we are able to get a fairly clean sample of binary systems with only a handful of false positives, with a precision of \todo{0.97} for the binary class. However, we do have a lot of false negatives, with only a completeness of \todo{0.32} for the binary class. The overall classification accuracy equals \todo{0.65}. The completeness of the binary class prediction is plotted against the temperature of the \gls{MS} companion to the \gls{YSG} primary star. Figure \ref{fig:YSG_completeness_companion_temperature} shows that the completeness increases with the companion temperature, as expected. The completeness is near 0.0 when the companion temperature is around 10,000 K (i.e., cool B-type stars), but rises up to $\sim0.7$ as the temperature gets closer to 50,000 K (i.e., hot O-type stars). A similar result was found in \citet{ogrady_2024}, who found that, using optical photometry, it is easier to identify \gls{YSG} binaries if the companion is hotter and more massive. However, they also found that including ultraviolet photometry allows them to reliably identify \gls{YSG} binaries with smaller companions that have masses of $\sim$7$M_\odot$. Thus, adding bluer bands from other surveys will likely improve classification accuracy here as well.

To summarize, we find that we can use \texttt{flexAE} to identify binary systems with a \gls{YSG} and O- or B-type \gls{MS} companion with Rubin \gls{LSST} at distances between \todo{500} - \todo{2,000} kpc. It is more challenging than finding \gls{MS}+\gls{MS} binaries with the Rubin \gls{LSST} bands, but it is still possible to find a pure sample of \gls{YSG} candidates. We note that we expect the model to improve if the model parameters, such as the number of latent space dimensions, are finetuned more carefully. We will use this model to find \gls{YSG} binary candidates in Rubin \gls{DP2} in future work.

\section{Size of Latent Space}
\label{app:size_latent_space}

One of the key decisions to make when working with \glspl{AE} is the number of latent space dimensions, $n_{z}$, to include in the model. This is a balance between interpretability and the quality of the reconstruction. If $n_{z}$ is too low, the \gls{AE} cannot retain enough information in the latent space to ensure reliable reconstruction. In contrast, if $n_{z}$ is too high, it becomes harder to interpret the latent space and the model takes longer to train. Typically, we aim to keep the number of latent space dimensions small, without compromising the reconstruction or \gls{BCE} loss.

\begin{figure}
	\includegraphics[width=0.8\columnwidth]{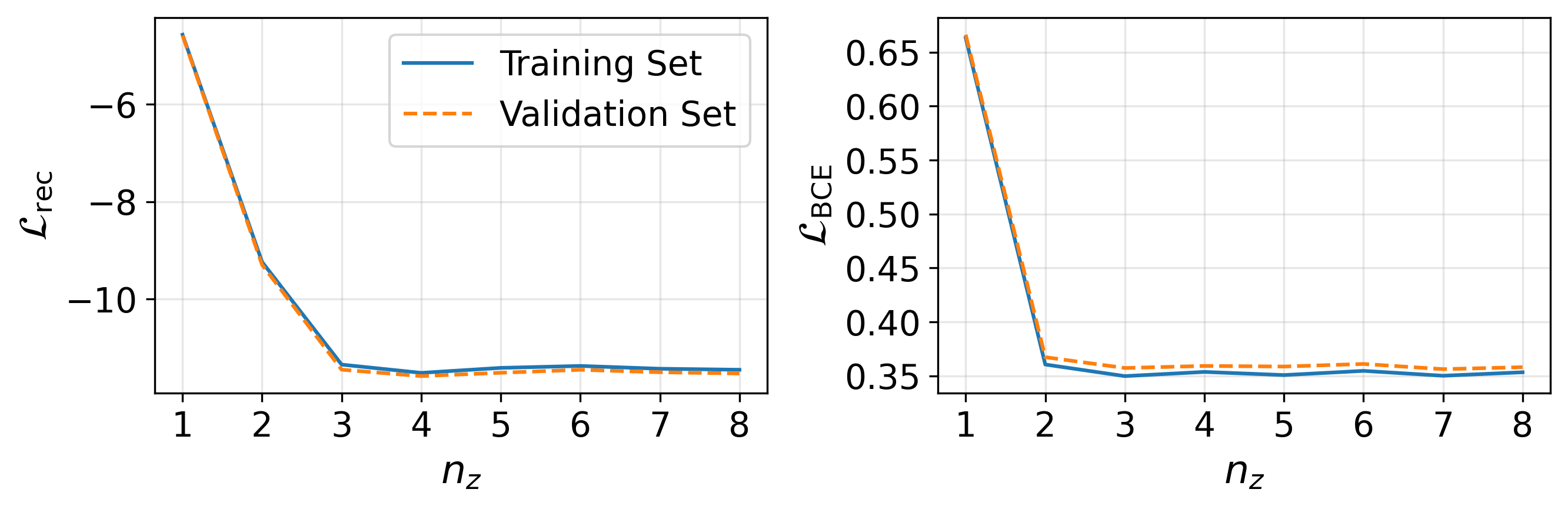}
    \centering
    \caption{The reconstruction loss ($\mathcal{L}_{\rm rec}$; left panel) and \gls{BCE} loss ($\mathcal{L}_{\rm BCE}$; right panel) for the simulated training set (blue) and simulated validation set (orange) as a function of the number of latent space dimensions ($n_{z}$). The models are otherwise identical to the main model used in this work (see Section \ref{sec:ae_architecture} and \ref{sec:training}). We want to keep the value of $n_{z}$ low, without compromising $\mathcal{L}_{\rm rec}$ or $\mathcal{L}_{\rm BCE}$. Thus, we choose $n_{z} = \todo{4}$ in this work.}
    \label{fig:latent_space_size}
\end{figure}

To find the ideal value of $n_{z}$, we train a number of \gls{AE} models with varying latent space dimensions. The \gls{AE} models are otherwise identical to the main model used in this work (see Sections \ref{sec:ae_architecture} and \ref{sec:training}). The reconstruction loss ($\mathcal{L}_{\rm rec}$) and \gls{BCE} loss ($\mathcal{L}_{\rm BCE}$) for these models are shown in the left and right panels of Figure \ref{fig:latent_space_size}, respectively. As expected, both $\mathcal{L}_{\rm rec}$ and $\mathcal{L}_{\rm BCE}$ are highest when $n_{z} = 1$. The value of $\mathcal{L}_{\rm BCE}$ decreases until $n_{z} = \todo{2}$, after which it plateaus. This suggests that only two latent space dimensions are sufficient for accurate classification in the simulated training and validation sets. The value of $\mathcal{L}_{\rm rec}$ plateaus after $n_{z} = \todo{3}$, which suggests that only three latent space dimensions are needed to capture all relevant information to reconstruct the input features. Improved reconstruction increases the chance that the model will generalize better to real data, suggesting that we should at least use three latent space dimensions. However, we include one more latent space dimension as a buffer and opt to use $n_{z} = \todo{4}$ in this work.

\section{Different model sizes}
\label{app:model_size}

Section \ref{sec:ae_architecture} and Figure \ref{fig:architecture} describe the model architecture used in this work. The model has multiple distinct components: the input heads,  encoder, decoder, latent space, and classifier. Here, we keep the general model architecture shown in Figure \ref{fig:architecture}, but change the size and shape of each of the individual components. We train and test \todo{five} different models that incrementally increase in complexity, named \emph{XS}, \emph{S}, \emph{M}, \emph{L}, and \emph{XL}. The sizes of each component for each model are detailed in Table \ref{tab:model_sizes}. Note that the number of trainable parameters also increases dramatically as we move from the \emph{XS} to the \emph{XL} model. 

\begin{figure}
	\includegraphics[width=0.8\columnwidth]{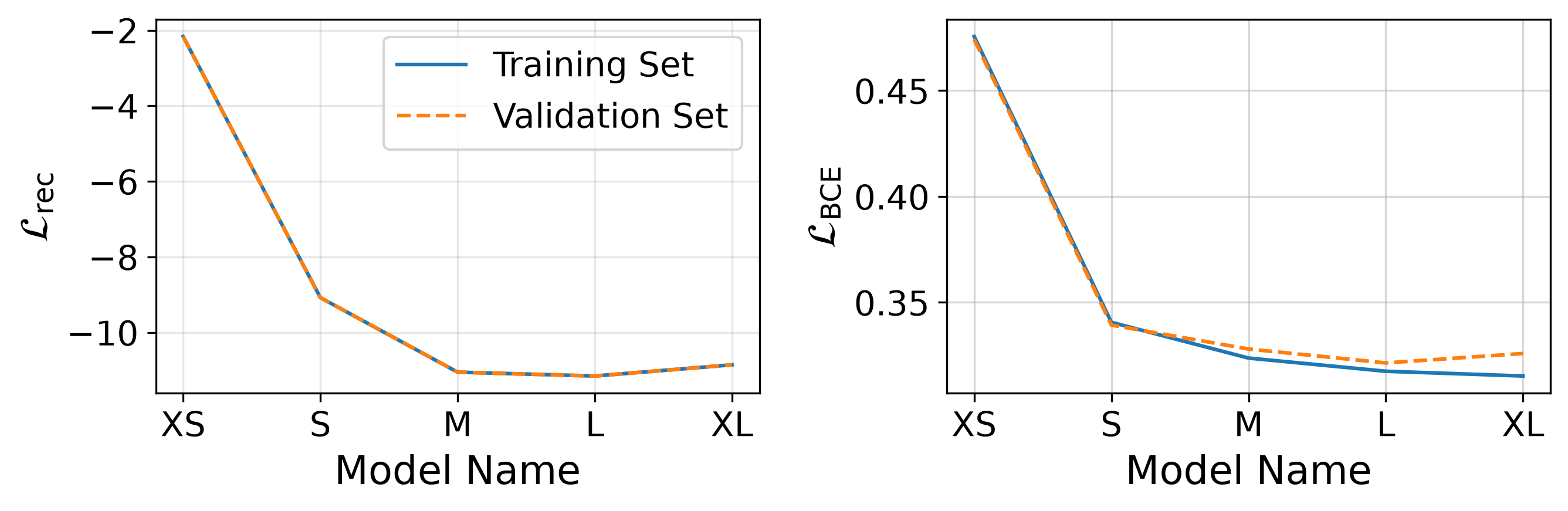}
    \centering
    \caption{The reconstruction loss ($\mathcal{L}_{\rm rec}$; left panel) and \gls{BCE} loss ($\mathcal{L}_{\rm BCE}$; right panel) for the simulated training set (blue) and simulated validation set (orange) for different models whose components have varying sizes and shapes (see Table \ref{tab:model_sizes}).}
    \label{fig:finetune_model_size}
\end{figure}

The effect of the model size on the reconstruction loss ($\mathcal{L}_{\rm rec}$) and \gls{BCE} loss ($\mathcal{L}_{\rm BCE}$) is shown in Figure \ref{fig:finetune_model_size}. We see that both $\mathcal{L}_{\rm rec}$ and $\mathcal{L}_{\rm BCE}$ decrease as we move from the \emph{XS} to the \emph{M} model, as the latter is more complex with more trainable parameters. However, the model performance stagnates as we consider models more complex than model \emph{M}. The improvements are only marginal, while the number of trainable parameters (and the time needed to train the models) keep increasing significantly. Thus, we decide to work with model \emph{M} in this work.

\begin{deluxetable*}{c c c c c c}
  \tablecaption{The details of the size and shape of the different components of the models tested. This includes the size and dimensionality of the input heads ($n_{\textrm{head}}$), the encoder ($n_{\textrm{encoder}}$), the latent space ($n_{\textrm{z}}$), the classifier ($n_{\textrm{classifier}}$), and the total number of trainable parameters ($n_{\textrm{train}}$). For example, if $n_{\textrm{encoder}}$ equals (16, 8), then the encoder has two layers with 16 and 8 output dimensions, respectively. We note that $n_{\textrm{head}}$ represents the size and shape of one input head, while the full model has two separate input heads (i.e., one head for the data and one for their uncertainties; see Figure \ref{fig:architecture}). Thus, the full size of the heads combined is twice $n_{\textrm{head}}$. The decoder component has the same size and dimensionality as the encoder, but reversed.}
  \label{tab:model_sizes}
  \tablehead{
    \colhead{\textbf{Model Name}} & \textbf{$n_{\textrm{head}}$} & \textbf{$n_{\textrm{encoder}}$} & \textbf{$n_{\textrm{z}}$} & \textbf{$n_{\textrm{classifier}}$} & \textbf{$n_{\textrm{train}}$}\\
    }
  \decimalcolnumbers
  \startdata
  \emph{XS} & 2 & 4 & 1 & 4 & 100\\
  \emph{S} & 6 & (8, 4) & 2 & (8, 8) & 461\\
  \emph{M} & 12 & (16, 8) & 4 & (16, 16, 16) & 1,779\\
  \emph{L} & (24, 24) & (32, 16) & 8 & (32, 32, 32) & 7,631\\
  \emph{XL} & (48, 48) & (128, 64, 32) & 16 & (64, 64, 64, 64) & 64,727\\
  \enddata
\end{deluxetable*}
\onecolumngrid

\section{Finetuning $\gamma$}
\label{app:finetune_gamma}

As shown in Equation \ref{eq:total_loss} in Section \ref{sec:loss_function}, the loss function is a combination of two separate losses: the reconstruction loss ($\mathcal{L}_{\rm rec}$) and the \gls{BCE} loss ($\mathcal{L}_{\rm BCE}$). However, both losses have very different scales. In principle, $\mathcal{L}_{\rm rec}$ can go up to +$\infty$ for extremely incorrect reconstructions. It can also become negative due to the penalty term. Meanwhile, a random classifier will have a $\mathcal{L}_{\rm BCE}$ of $\sim$0.693, which can go down to 0 for a perfect classifier. We added the proportionality constant $\gamma$ in Equation \ref{eq:total_loss} to balance these two loss terms with different scales.

\begin{figure}
	\includegraphics[width=0.8\columnwidth]{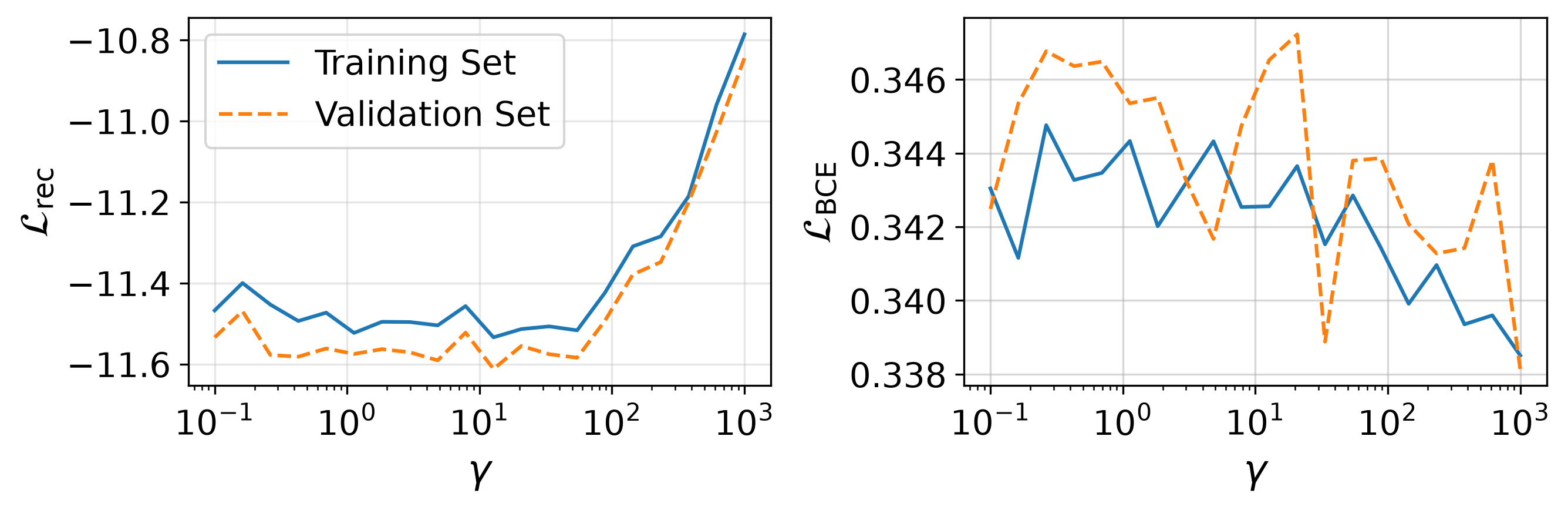}
    \centering
    \caption{The reconstruction loss ($\mathcal{L}_{\rm rec}$; left panel) and \gls{BCE} loss ($\mathcal{L}_{\rm BCE}$; right panel) for the simulated training set (blue) and simulated validation set (orange) as a function of the value of the proportionality constant between the two losses, $\gamma$ (see Equation \ref{eq:total_loss}). The models are otherwise identical to the main model used in this work (see Section \ref{sec:ae_architecture} and \ref{sec:training}). As expected, $\mathcal{L}_{\rm BCE}$ goes down as $\gamma$ increases, while $\mathcal{L}_{\rm rec}$ goes up.}
    \label{fig:finetune_gamma}
\end{figure}

Choosing the right value of $\gamma$ is a trade-off between both losses. The end goal is accurate classification, so minimizing $\mathcal{L}_{\rm BCE}$ is crucial. However, we need to keep an eye on $\mathcal{L}_{\rm rec}$ as well, as improved reconstruction increases the chance that the model will generalize better from simulated data to real data. To find the ideal value of $\gamma$, we train a number of \gls{AE} models with different values of $\gamma$ in the loss function. The architecture and training procedure of the models are otherwise identical to what was described in the main text (see Sections \ref{sec:ae_architecture} and \ref{sec:training}). The values of $\mathcal{L}_{\rm rec}$ and $\mathcal{L}_{\rm BCE}$ for these models are shown in the left and right panels of Figure \ref{fig:finetune_gamma}, respectively. We see the expected tug-of-war behavior: $\mathcal{L}_{\rm rec}$ goes up as $\gamma$ increases, while $\mathcal{L}_{\rm BCE}$ decreases. However, it is worth noting that the difference in both losses is actually minimal for the whole range of $\gamma$ tested. The value of $\mathcal{L}_{\rm rec}$ only varies between \todo{-11.6} to \todo{-10.8}, while $\mathcal{L}_{\rm BCE}$ varies between \todo{0.338} and \todo{0.346}. We aim to choose the value of $\gamma$ that results in low values of $\mathcal{L}_{\rm BCE}$ without compromising reconstruction. Thus, we choose to use $\gamma = \todo{20}$ for this work.

\section{Training set size}
\label{app:training_set_size}

A major advantage of training the \gls{AE} with a simulated dataset is that the simulated training set can be made arbitrarily large. However, the model will take longer to train if the training set is larger. We want to ensure that the simulated training set is large enough so that the \gls{AE} is able to learn the full diversity of single stars and binary systems, and capture this diversity in its latent space. 

\begin{figure}
	\includegraphics[width=0.8\columnwidth]{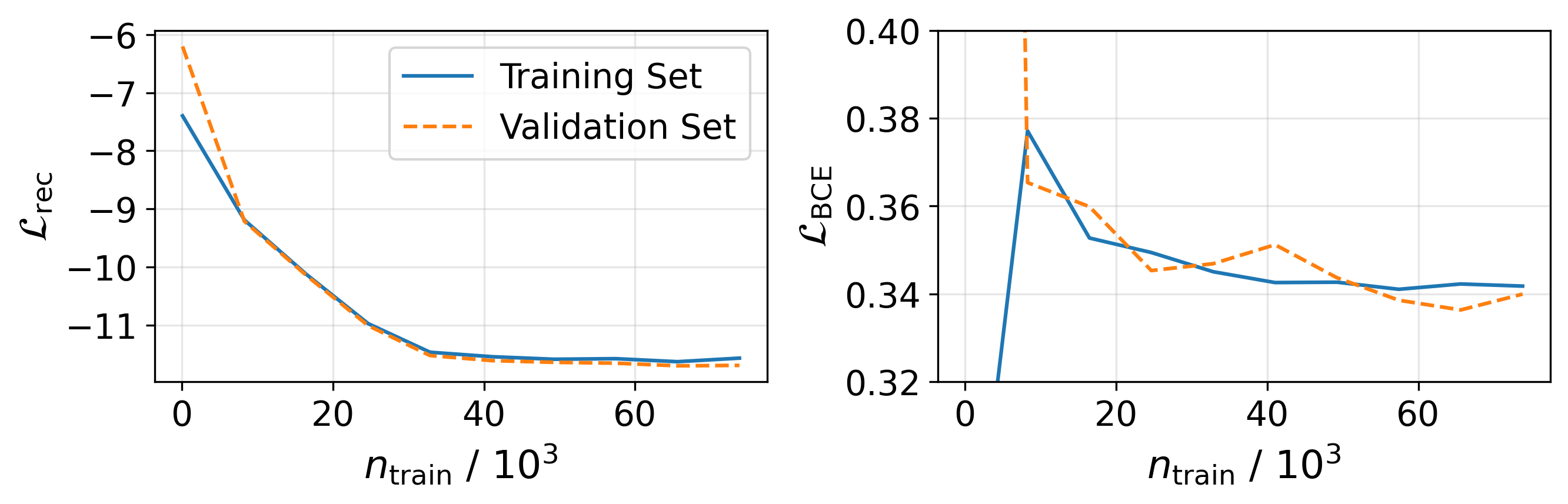}
    \centering
    \caption{The reconstruction loss ($\mathcal{L}_{\rm rec}$; left panel) and \gls{BCE} loss ($\mathcal{L}_{\rm BCE}$; right panel) for the simulated training set (blue) and simulated validation set (orange) as a function of the size of the simulated training set ($n_{\rm train}$). Both $\mathcal{L}_{\rm rec}$ and $\mathcal{L}_{\rm BCE}$ decrease as the sample size increases, and plateau after a sample size of \todo{40,000} targets.}
    \label{fig:finetune_samplesize}
\end{figure}

We train a number of \gls{AE} models, as described in Sections \ref{sec:ae_architecture} and \ref{sec:training}, but vary the size of the simulated training set in each iteration. We aim to find a training set size where the reconstruction loss ($\mathcal{L}_{\rm rec}$) and the \gls{BCE} loss ($\mathcal{L}_{\rm BCE}$) plateau, so that we are not missing out on model improvements by not increasing the training set size. The change in $\mathcal{L}_{\rm rec}$ and $\mathcal{L}_{\rm BCE}$ with respect to the size of the training set is shown in Figure \ref{fig:finetune_samplesize}. As expected, both $\mathcal{L}_{\rm rec}$ and $\mathcal{L}_{\rm BCE}$ decrease as the sample size increases. Note that the value of $\mathcal{L}_{\rm BCE}$ for the training set for small sample sizes ($<10,000$) is very low, while the validation loss is very high. This is a classic sign of overfitting. Both losses plateau after a simulated training set size of \todo{40,000}, which is what we use in the main part of this work.

\section{Differences in the distribution of $\hat{y}$}
\label{app:yhat_distribution}

\begin{figure}
	\includegraphics[width=0.5\columnwidth]{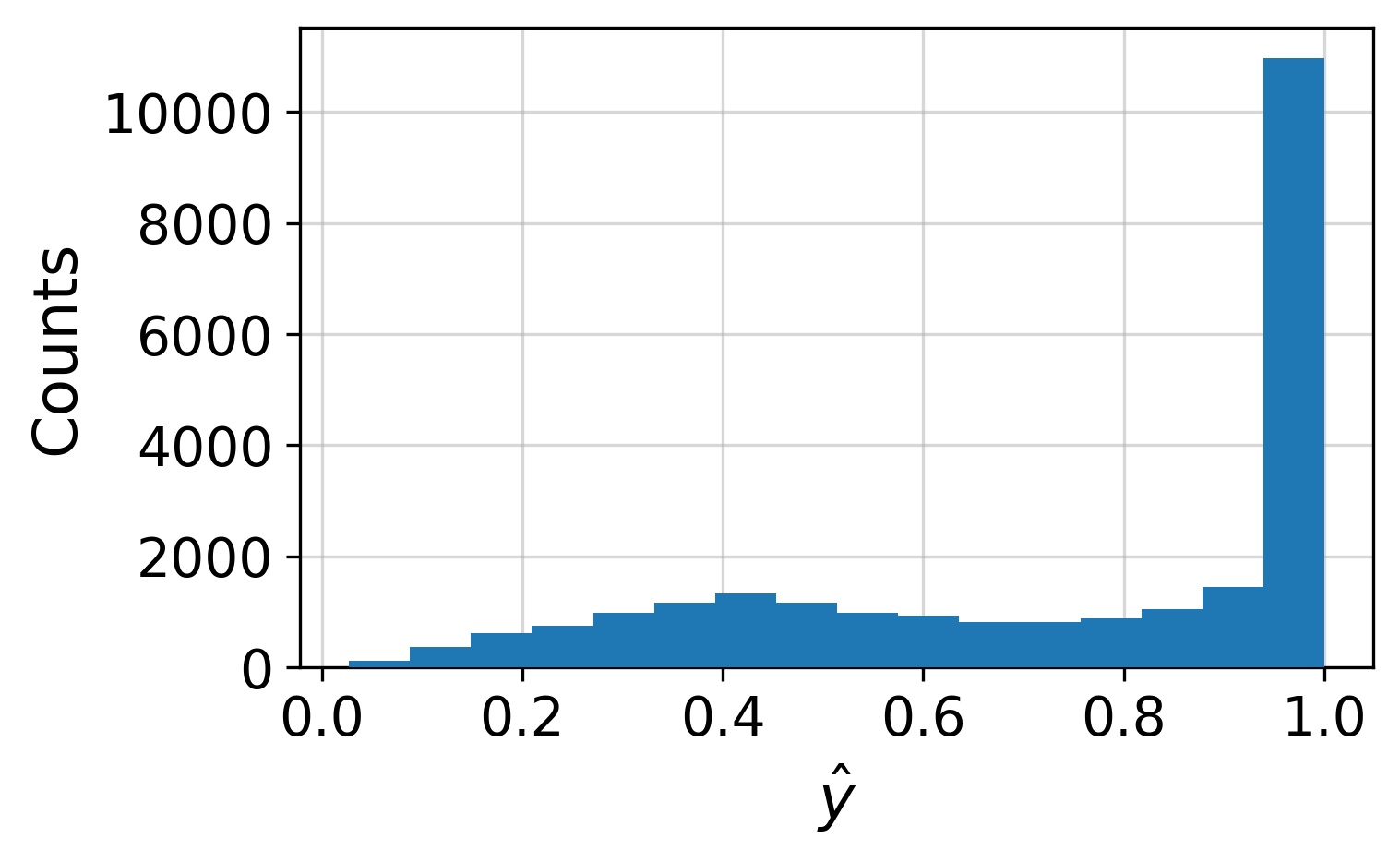}
    \centering
    \caption{The model class predictions ($\hat{y}$) for the test set after explicitly removing all the simulated targets that appear on the blue side of the 47 Tucanae isochrone.}
    \label{fig:yhat_noblueintraining}
\end{figure}

In Section \ref{sec:dp1_binary_47tuc}, we noted that the distribution of the class predictions ($\hat{y}$) differed between the simulated test set (see bottom-right panel of Figure \ref{fig:classification}) and the \gls{DP1} set (see Figure \ref{fig:yhat_dp1}). We argue that this difference can be best interpreted as differences in the relative proportions of the types of targets in the simulated set and the \gls{DP1} set. Most notably, the first peak in $\hat{y}$ moved from $\hat{y} \sim 0$ for the simulated set to $\hat{y} \sim 0.3$ for the \gls{DP1} set. This is because we varied the luminosity by $\pm$0.2 dex while generating the simulated set to increase the diversity of training data (see Section \ref{sec:creating_training_set}). This caused a lot of simulated targets to appear on the blue side of the 47 Tucanae isochrone in the $i$ vs. $g-i$ \gls{CMD}, which are confidently classified as single stars with low values of $\hat{y}$. In contrast, the real \gls{DP1} dataset does not have a lot of these targets, and so we do not expect a large peak at $\hat{y} \sim0$ in \gls{DP1} set. Indeed, Figure \ref{fig:yhat_noblueintraining} shows that after explicitly removing all targets from the simulated set that appear on the blue side of the 47 Tucanae isochrone, the first peak in the distribution of $\hat{y}$ for the simulated set disappears entirely, as expected, while the targets with $\hat{y} \sim0.3-0.4$ become more apparent. Note that the peak at $\hat{y} \sim 1.0$ is still larger in Figure \ref{fig:yhat_noblueintraining} compared to the distribution in Figure \ref{fig:yhat_dp1}. This is because we have not adjusted the relative proportions of single stars and binary systems in the simulated set. Both appear equally frequently in the simulated set, while this is not the case in the real \gls{DP1} data.

\bibliography{bibtex}{}
\bibliographystyle{aasjournal}



\end{document}


%% file: glossary.tex
%
%
%
%
%

\newacronym{LSST}{LSST}{Legacy Survey of Space and Time}
\newacronym{JWST}{JWST}{James Webb Space Telescope}
\newacronym{HST}{HST}{Hubble Space Telescope}
\newacronym{GZ}{GZ}{Galaxy Zoo}
\newacronym{IFU}{IFU}{integral field unit}
\newacronym{MaNGA}{MaNGA}{Mapping Nearby Galaxies at Apache Point Observatory}
\newacronym{DynPop}{DynPop}{Dynamics and stellar Population}
\newacronym{CALIFA}{CALIFA}{Calar Alto Legacy Integral Field Area}
\newacronym{ALFALFA}{ALFALFA}{Arecibo Legacy Fast ALFA}
\newacronym{ATLAS}{ATLAS}{Asteroid Terrestrial-impact Last Alert System}
\newacronym{ALMA}{ALMA}{Atacama Large Millimeter/submillimeter Array}
\newacronym{GOTO}{GOTO}{Gravitational-wave Optical Transient Observer}
\newacronym{WFPC2}{WFPC2}{Wide Field Planetary Camera 2}
\newacronym{DECam}{DECam}{Dark Energy Camera}
\newacronym{ZTF}{ZTF}{Zwicky Transient Facility}
\newacronym{PTF}{PTF}{Palomar Transient Factory}
\newacronym{ACS}{ACS}{Advanced Camera for Surveys}
\newacronym{WFC}{WFC}{Wide Field Channel} 
\newacronym{DR1}{DR1}{Data Release 1}
\newacronym{DR2}{DR2}{Data Release 2}
\newacronym{DR3}{DR3}{Data Release 3}
\newacronym{DP2}{DP2}{Data Preview 2}
\newacronym{DP1}{DP1}{Data Preview 1}
\newacronym{DP0}{DP0}{Data Preview 0}
\newacronym{YSE}{YSE}{Young Supernova Experiment}
\newacronym{SDSS}{SDSS}{Sloan Digital Sky Survey}
\newacronym{TNS}{TNS}{Transient Name Server}
\newacronym{LSSTComCam}{LSSTComCam}{LSST Commissioning Camera}
\newacronym{LSSTCam}{LSSTCam}{LSST Camera}
\newacronym{DESC}{DESC}{Dark Energy Science Collaboration}
\newacronym{IAU}{IAU}{International Astronomical Union}
\newacronym{ATCA}{ATCA}{Australia Telescope Compact Array}
\newacronym{ALeRCE}{ALeRCE}{Automatic Learning for the Rapid Classification of Events}

\newacronym{DM}{DM}{dark matter}
\newacronym{BH}{BH}{black hole}
\newacronym{IMF}{IMF}{initial mass function}
\newacronym{AGN}{AGN}{active galactic nucleus}
\newacronym{M/L}{M/L}{mass-to-light ratio}
\newacronym{NEA}{NEA}{near-Earth asteroids}
\newacronym{RSG}{RSG}{red supergiant}
\newacronym{YSG}{YSG}{yellow supergiant}
\newacronym{MIR}{MIR}{mid-infrared}
\newacronym{SED}{SED}{spectral energy distribution}
\newacronym{CSM}{CSM}{circumstellar material}
\newacronym{TDE}{TDE}{tidal disruption event}
\newacronym{LBV}{LBV}{luminous blue variable}
\newacronym[shortplural=SN Type Ia,longplural=Type Ia supernovae]{SN Type Ia}{SN Type Ia}{Type Ia supernova}
\newacronym[shortplural=SNe,longplural=supernovae]{SN}{SN}{supernova}
\newacronym{MS}{MS}{main sequence}
\newacronym{CMD}{CMD}{color-magnitude diagram}
\newacronym{WD}{WD}{white dwarf}
\newacronym{ODF}{ODF}{opacity distribution function}
\newacronym{SMC}{SMC}{Small Magellanic Cloud}
\newacronym{LMC}{LMC}{Large Magellanic Cloud}
\newacronym{IMBH}{IMBH}{intermediate-mass black hole}

\newacronym{SNR}{SNR}{signal-to-noise ratio}
\newacronym{BCE}{BCE}{binary cross-entropy}
\newacronym{AE}{AE}{autoencoder}
\newacronym{VAE}{VAE}{variational autoencoder}
\newacronym{MSE}{MSE}{mean squared error}
\newacronym{KLD}{KLD}{Kullback–Leibler divergence}
\newacronym{TW}{TW}{Tremaine-Weinberg}
\newacronym{JAM}{JAM}{Jeans anisotropic modeling}
\newacronym{MGE}{MGE}{Multi-Gaussian Expansion}
\newacronym{SPS}{SPS}{stellar population synthesis}
\newacronym{FWHM}{FWHM}{full width at half maximum}
\newacronym{PSF}{PSF}{point spread function}
\newacronym{ATClean}{\texttt{ATClean}}{ATLAS Clean}
\newacronym{CHIPS}{CHIPS}{Complete History of Interaction-Powered Supernovae}
\newacronym{DETECT}{\texttt{DETECT}}{Detection Efficiency and Threshold Estimation for Characterization of Transients}
\newacronym{RSP}{RSP}{Rubin Science Platform}
\newacronym{DC2}{DC2}{Data Challenge 2}
\newacronym{ML}{ML}{machine learning}
\newacronym{SVM}{SVM}{support vector machine}
\newacronym{MLP}{MLP}{multi-layer perceptron}
\newacronym{NN}{NN}{neural network}
\newacronym{GELU}{GELU}{Gaussian Error Linear Unit}